\documentclass[usenatbib]{mnras}

\usepackage{graphicx}	% Including figure files
\usepackage{subfigure}
\usepackage{amsmath}	% Advanced maths commands
\usepackage{amssymb}	% Extra maths symbols
\usepackage{multicol}        % Multi-column entries in tables
\usepackage{bm}		% Bold maths symbols, including upright Greek
\usepackage{pdflscape}	% Landscape pages
\usepackage{threeparttable}
\usepackage{booktabs}

\usepackage[T1]{fontenc}
\usepackage{ae,aecompl}

\usepackage{newtxtext,newtxmath}
\title[Probing cosmic background dynamics]{\textit{Probing Cosmic Background Dynamics with a Cosmological-model-independent Method}}%Probe cosmic background dynamics with a cosmological-model-independent method

\author[Y. Liu et al.]{Yang Liu,$^{1}$\thanks{yangl@hunnu.edu.cn} Bao Wang,$^{2,3}$\thanks{baowang@pmo.ac.cn} Hongwei Yu$^{1,4}$\thanks{Corresponding author: hwyu@hunnu.edu.cn} and Puxun Wu$^{1,4}$\thanks{Corresponding author: pxwu@hunnu.edu.cn}
	\\
	$^{1}$Department of Physics and Synergetic Innovation Center for Quantum Effects and Applications, Hunan Normal University, Changsha, Hunan 410081, China\\
	$^{2}$School of Astronomy and Space Sciences, University of Science and Technology of China, Hefei 230026, China\\
	$^{3}$Purple Mountain Observatory, Chinese Academy of Sciences, Nanjing 210023, China\\
	$^{4}$Institute of Interdisciplinary Studies, Hunan Normal University, Changsha, Hunan 410081, China	
}	

\pubyear{2024}

\begin{document}
	\label{firstpage}
	\pagerange{\pageref{firstpage}--\pageref{lastpage}}
	\maketitle

\begin{abstract}

The Hubble constant $H_0$ tension has emerged as the most serious crisis in modern cosmology, potentially indicating that the $\Lambda$CDM model may not describe our universe accurately. In this paper, we establish a new, cosmological-model-independent method to study the cosmic background dynamics. 
Using the latest Pantheon+ Type Ia supernova (SN Ia) sample and the model-independent SN Ia sample (P+1690), we derive values for the luminosity distance, the Hubble parameter, and the deceleration parameter at five different redshift points ranging from 0.12 to 0.52. Our analysis shows that results obtained from the Pantheon+ sample align with the predictions of  the $\Lambda$CDM model within 2$\sigma$ confidence level (CL),  while those obtained from the P+1690 sample exhibit deviations of about $2\sim3\sigma$ CL.
Furthermore, we explore the equation of state (EoS) of dark energy and find that while the EoS values from the Pantheon+ sample remain consistent with  $-1$  within 2$\sigma$ CL, the P+1690 sample does not conform to this standard.
These findings remain unchanged after  the inclusion of the Hubble parameter measurements in our analysis. 
Our results indicate that the $\Lambda$CDM model remains compatible with the Pantheon+ SN Ia and the Hubble parameter measurements at 2$\sigma$  CL.

\end{abstract}

\begin{keywords}
	\emph{(cosmology:)} cosmological parameters -- cosmology: observations
\end{keywords}

%%%%%%%%%%%%%%%%%%%%%%%%%%%%%%%%%%%%%%%%%%%%%%%%%%

%%%%%%%%%%%%%%%%% BODY OF PAPER %%%%%%%%%%%%%%%%%%
	
\section{Introduction}
The cosmological constant $\Lambda$ plus cold dark matter ($\Lambda$CDM) is the simplest and  most  favored cosmological model  to describe the cosmic evolution. The $\Lambda$CDM model, although fits observational data very well, still faces some challenges.  Among them, the Hubble constant ($H_0$) tension is the most notable one and it has been considered as  the most serious crisis in modern cosmology~\citep{Riess2020,Perivolaropoulos2022,Tully2023,Liu2023,Liu2024}.  The $H_0$ tension  refers to the discrepancy  (more than 5$\sigma$) between the measurements of $H_0$ using the near type Ia supernova (SN Ia) calibrated by Cepheids \citep{Riess2022} and that from the high-redshift cosmic microwave background (CMB) radiation observation within the framework of  the $\Lambda$CDM model~\citep{Planck2020}, and it indicates that the assumed $\Lambda$CDM model used to determine the Hubble constant may be inconsistent with our present Universe or there may be potentially unknown systematic errors in the observational data. It is worth noting, however, that any systematics, which  could explain the $H_0$ tension, have not been found~\citep{Efstathiou2014,Feeney2018,Riess2016,Cardona2017,Zhang2017,Follin2018,Riess2018a,Riess2018b}.
Thus, it is necessary to investigate whether the $\Lambda$CDM model can  correctly  describe our Universe.

The cosmological constant as dark energy has a constant equation of state (EoS) parameter $w_\mathrm{DE}$ equal to $-1$, where $w_\mathrm{DE}$ is defined as $w_\mathrm{DE} \equiv \frac{P_\mathrm{DE}}{\rho_\mathrm{DE}}$ with $P_\mathrm{DE}$ and $\rho_\mathrm{DE}$ being  the pressure and energy density of dark energy, respectively. 
Thus, generalizing the EoS of dark energy from $-1$ to  an arbitrary constant $w_\mathrm{DE}$ or a parametrized form as a function of redshift $z$, 
e.g., the Chevalier-Polarski-Linder (CPL) parameterization~\citep{Chevallier2001,Linder2003} with $w_\mathrm{DE}(z)=w_0+w_az/(1+z)$, and constraining $w_\mathrm{DE}$ or the coefficients in the parameterization from observational data, we can judge the viability  of  the cosmological constant as dark energy by analyzing whether $w_\mathrm{DE}=-1$ is allowed by the observations~\citep{Liu2008,Demianski2020}. 
Recently, by combining the baryon acoustic oscillation from the first year of observations from the Dark Energy Spectroscopic Instrument (DESI)~\citep{DESI2024} with CMB anisotropies from Planck~\citep{Planck2020,Planck2020b} and CMB lensing data from Planck and Atacama Cosmology Telescope~\citep{Carron2022,Qu2024}, and DES-SN5YR supernova datasets~\citep{Abbott2024}, \cite{DESI2024} found that the time-varying dark energy EoS parametrized by the CPL model is more favored than $w_\mathrm{DE}=-1$ with a statistical significance of $3.9\sigma$.

Reconstructing the cosmic background evolution directly from the observations is a more reliable method to understand the expanding history of our Universe.  The usual methods include  the nonparametric Bayesian reconstruction~\citep{Zhao2012,Zhao2017} and the Gaussian process~\citep{Holsclaw2010,Seikel2012,Shafieloo2012}.
Using the observational data, e.g., SN Ia, one can reconstruct the Hubble parameter ($H(z)$) or the luminosity distance ($d_L(z)$) with their derivatives, and then compare them with the predictions from the $\Lambda$CDM model  to determine whether the $\Lambda$CDM model can correctly describe  the cosmic evolution. 
However, when reconstructing cosmic evolution in low (high) redshift regions, high (low) redshift observational data are utilized concurrently. Consequently, the low (high) redshift data influence the reconstructed results for the high (low) redshift regions.

In this work, we establish  a new method to obtain  the cosmic background  dynamics in different redshift regions from observational data. 
In our method, we  only assume the cosmological principle, which implies that the universe can be described using the Friedmann-Lema\^{i}tre-Robertson-Walker~(FLRW) metric. Therefore, our approach is metric-dependent. However, it does not necessitate the assumption of any specific energy components within the universe, and in this sense,  our method is cosmological-model-independent.
Furthermore, when the properties of  the cosmic expansion are studied at a given redshift, the observational data in the near region of this redshift point rather than the full data will be used.  
From the Pantheon+ SN Ia and the Hubble parameter measurements, we obtain the values of the Hubble parameter and the deceleration parameter at different redshifts, and find that they are consistent with  the predictions of the $\Lambda$CDM model at $2\sigma$ confidence level (CL).
	
\section{Method}\label{sec:2}
For a homogeneous and isotropic Universe described  by the FLRW metric, the Hubble parameter $H$, which gives the cosmic expanding velocity, is defined as 
\begin{eqnarray}
	H\equiv \frac{1}{a}\frac{da}{dt}\,,\label{eq:H}
\end{eqnarray} 
where $a$ is the cosmic scale factor and $t$ the cosmic time. 
In a spatially flat $\Lambda$CDM model, the Hubble parameter has the form: $H(z)=H_0\sqrt{\Omega_\mathrm{m0}(1+z)^3+(1-\Omega_\mathrm{m0})}$ with $\Omega_\mathrm{m0}$ being the present matter density parameter. 
Using the Hubble parameter, one can obtain the luminosity distance $d_L(z)$:
\begin{eqnarray}
	d_L(z)=(1+z)\int_{0}^z \frac{1}{H(z)}\, dz
\end{eqnarray}
in a spatially flat universe, where the velocity of light is set to  1.
Comparing the theoretical value and the observational one of the luminosity distance can yield  constraints on the cosmological parameters, i.e. $\Omega_\mathrm{m0}$, after choosing a concrete cosmological  model.  To cosmological-model-independently understand the cosmic dynamics, we perform the Taylor expansion of the luminosity distance at a given redshift $z_i$ and then obtain:
\begin{align}\label{eq:dl}
	d_L(z)
	&=d_{L,i}+(z-z_i)
	\left(\frac{1+z_i}{H_i}+\frac{d_{L,i}}{1+z_i}\right)\nonumber\\
	& +(z-z_i)^2\left(\frac{1}{H_i}-\frac{1+q_i}{2 H_i} \right)+\mathcal{O}\left((z-z_i)^3\right) \, ,
\end{align}
where $H_i=H(z_i)$, $d_{L,i}=d_L(z_i)$, and $q_i=q(z_i)$ with $q\equiv-\frac{1}{aH^2} \frac{d^2a}{dt^2}$ being the cosmic deceleration parameter, are  three free parameters. If we can determine their values from the observational data, the cosmic dynamics will be known. Since the convergence region of the Taylor series of the luminosity distance and the Hubble parameter  is the near region around $z = z_i$, we only consider the observational data in the redshift region $|z-z_i|\leq \Delta z$ to constrain $d_{L,i}$, $H_i$ and $q_i$,  where $\Delta z$ represents the convergence radius. When $z_i=0$, our method reduces to the usual cosmographic one, which has been widely used to study the cosmic expanding history~\citep{Visser2005,Luongo2011,Aviles2012,Dunsby2016,Capozziello2019,Capozziello2020,Mehrabi2021,Gao2023,Zhang2023}.  

Once the constraints on $H_i$ and $q_i$ at a given redshift are obtained, we can  calculate the EoS parameter of dark energy $w_{\mathrm{DE},i}$ at that redshift
\begin{eqnarray}\label{eq:w}
	w_{\mathrm{DE},i}= \frac{H_i^2(1-2q_i)}{3\left[H_0^2\Omega_\mathrm{m0}(1+z_i)^3-H_i^2\right]}
\end{eqnarray}
after assuming that the energy component of the Universe consists of pressureless matter and dark energy and the  Universe is spatially flat.
	
\section{Samples and Results}\label{sec:3}
\subsection{Pantheon+ SN Ia}
The  latest Pantheon+ SN Ia  sample~\citep{Scolnic2022} will be used firstly to constrain  parameters $d_{L,i}$, $H_i$, and $q_i$, which includes 1701 light-curves of 1550 distinct SN~Ia. 
The allowed regions for the parameters $d_{L,i}$, $H_i$, and $q_i$ can be determined using the Markov Chain Monte Carlo (MCMC) method to minimize the $\chi^2_\mathrm{SN}$, expressed as:
\begin{eqnarray}\label{eq:chi}
	\chi^2_\mathrm{SN}=\bm{\hat{Q}}^\dagger C_\mathrm{P+}^{-1} \bm{\hat{Q}}, 
\end{eqnarray}
where $\bm{\hat{Q}} \equiv \bm{\hat{\mu}}_\mathrm{obs}^\mathrm{corr}-\mu_\mathrm{th}$, with $\bm{\hat{\mu}}_\mathrm{obs}^\mathrm{corr}$ representing the array of observed corrected SN Ia distance modulus and $\mu_\mathrm{th}$ being the corresponding theoretical values, and $C_\mathrm{P+}$ is the covariance matrix.
The corrected SN~Ia distance modulus are obtained from~\citep{Tripp1998,Brout2022}
\begin{eqnarray}\label{eq:m_obs}
	\mu_\mathrm{obs}^\mathrm{corr} = m_B - M_B + \alpha x_1 - \beta c + \delta_\mathrm{host} - \delta_\mathrm{bias},
\end{eqnarray} 
where $m_B$ is  the apparent magnitude in the B-band filter, which  is related to the SALT2 light-curve  amplitude $x_0$ by $m_B=-2.5\log (x_0)$, $x_1$ is the stretch parameter, $c$ is the light-curve color, $\alpha$ and $\beta$ are coefficients relating luminosity to $x_1$ and $c$, respectively, $M_B$ is the fiducial B-band absolute magnitude of SN~Ia, $\delta_\mathrm{host}$ accounts for host-galaxy mass luminosity correction, and $\delta_\mathrm{bias}$ corrects for selection biases from simulations following~\citep{Brout2021,Popovic2021}.
The value of  $\mu_\mathrm{th}$   can be derived from the luminosity distance: 
\begin{eqnarray}\label{eq:mth}
	\mu_\mathrm{th}(z)=25+5\log\left(d_L(z)\right).
\end{eqnarray}
Given the high degeneracy between the Hubble parameter $H_i$ and $M_B$, a Gaussian prior of $-19.253\pm0.027~\mathrm{mag}$ is used for $M_B$ when only SN~Ia data are utilized. The covariance matrix in the Pantheon+ sample includes statistical ($C_\mathrm{stat}$) and systematic ($C_\mathrm{syst}$) components, addressing uncertainties from measurement errors, gravitational lensing, and peculiar-velocity effects. The redshift $z$ used is the Hubble-diagram redshift $z_\mathrm{HD}$, derived from the CMB frame redshift $z_\mathrm{CMB}$ with corrections for peculiar velocity.

For the Pantheon+ sample with $z_\mathrm{HD}$, we exclude those data whose redshifts are less than 0.01 since the unmodeled peculiar velocities will strongly impact the nearby SN Ia sample~\citep{Brout2022}.
Before using these real data to constrain the free parameters, we need to check the reliability of our method.  
To do so, we mock the SN Ia data from the fiducial model to assess the impact of different $\Delta z$ ( $\Delta z=0.15$ and $0.12$) on the results. The detailed discussions can be found in the Appendix \ref{appendix}. The results indicate that when  $\Delta z=0.15$ is used, the parameter $q_1$ is consistent with the fiducial model only at the margin of $1\sigma$ CL, whereas it conforms well with the model when using $\Delta z=0.12$.
Thus, $\Delta z=0.12$ is chosen when the Pantheon+ sample is utilized. We consider five expansion  points ($z_i$) in redshift from 0.12 to 0.52 with an increment of 0.1 in our analysis.
The number of SN Ia data in each redshift region is summarized in Table~\ref{tab1}.

\begin{figure*}
	\centering
	\includegraphics[width=0.66\columnwidth]{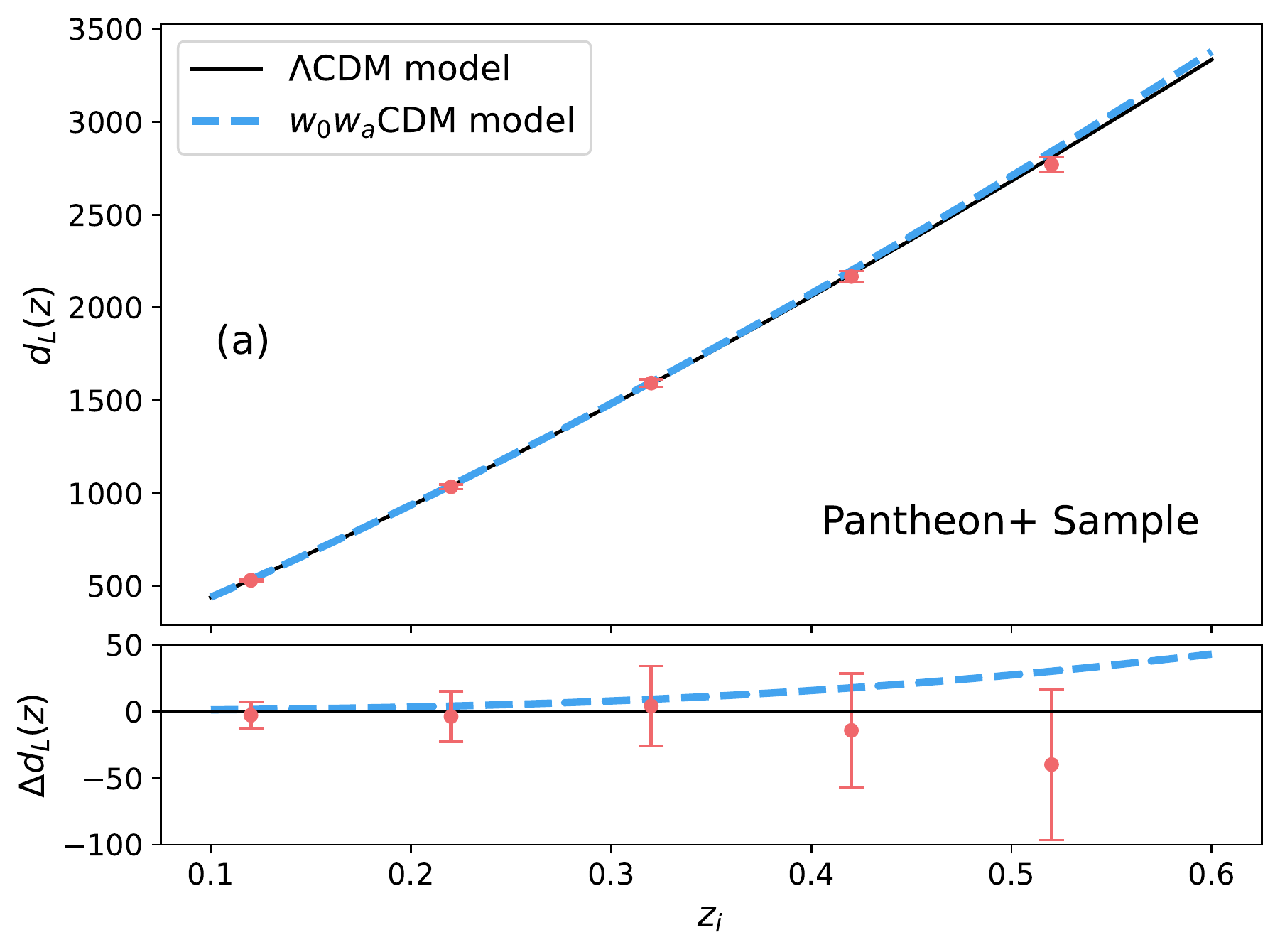}
	\includegraphics[width=0.66\columnwidth]{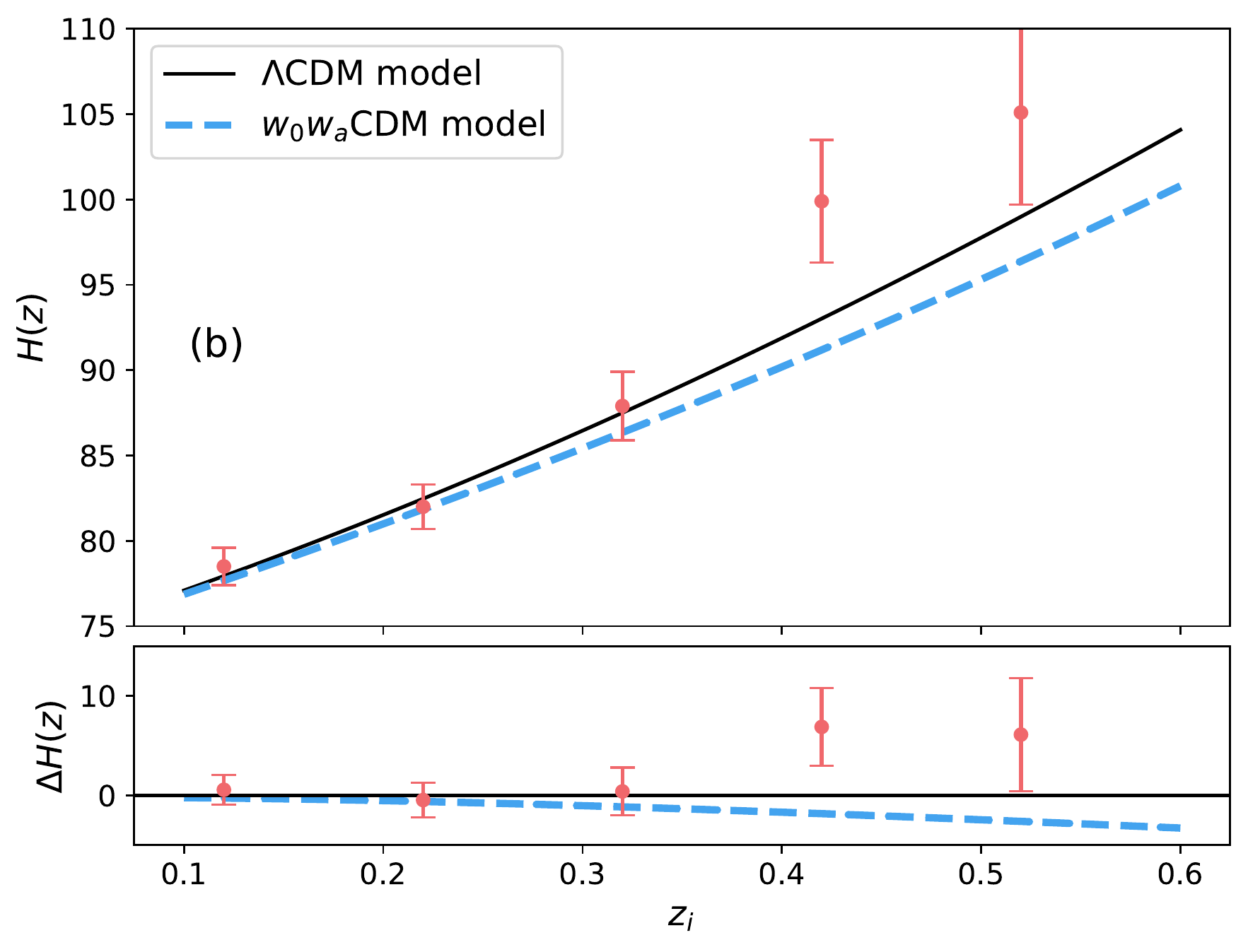}	
	\includegraphics[width=0.66\columnwidth]{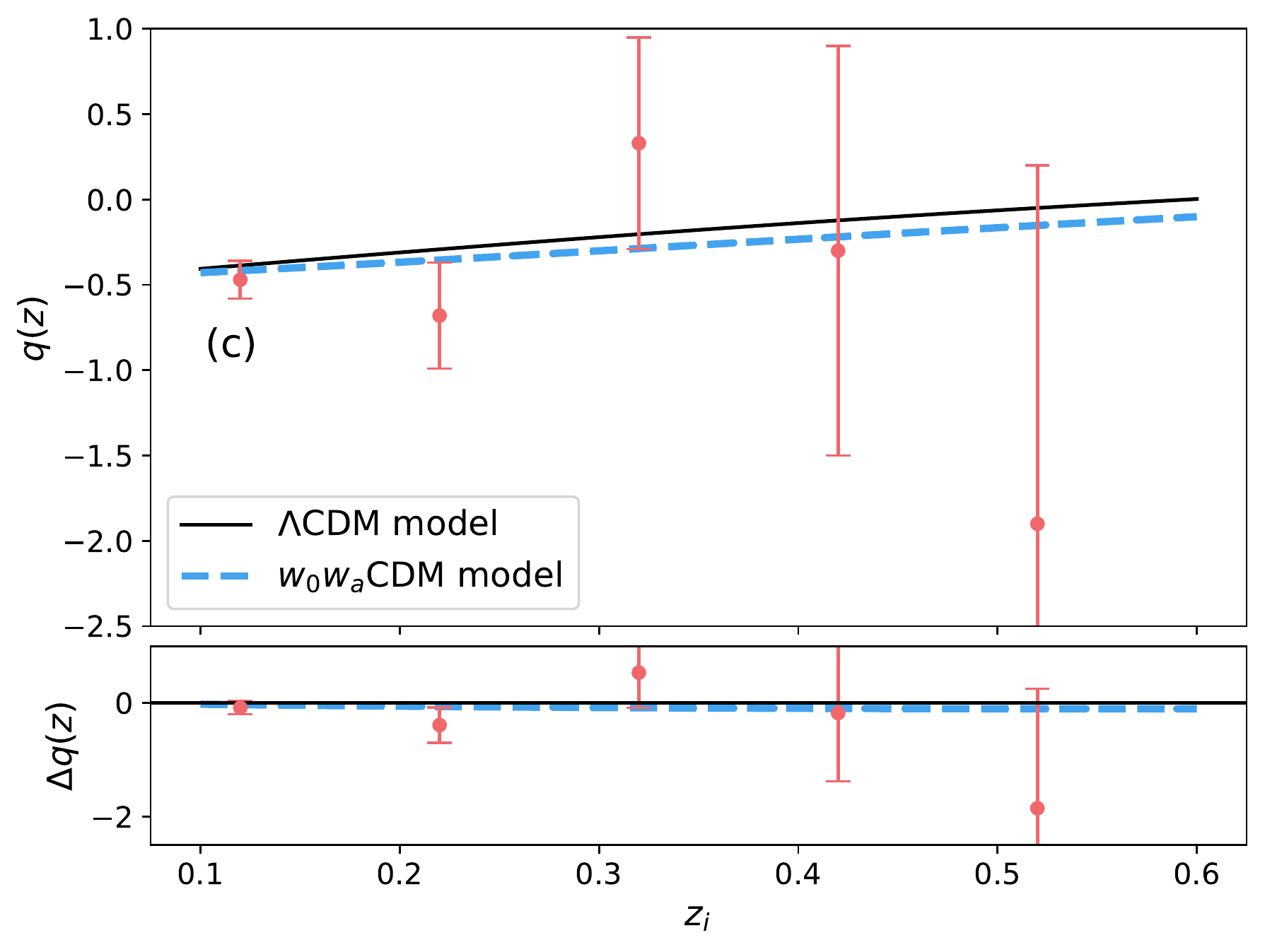}
	\caption{ The constraints on $d_{L,i}$, $H_i$ and $q_i$ at different redshift points by using the Pantheon+ sample. 
		The black solid and blue dashed lines represent, respectively, the predictions of the $\Lambda$CDM model with $H_0=73.2\pm0.94~\mathrm{km~s^{-1}Mpc^{-1}}$ and $\Omega_\mathrm{m0}=0.33\pm0.018$ and the $w_0w_a$CDM model with $H_0=72.98\pm0.94~\mathrm{km~s^{-1}Mpc^{-1}}$, $\Omega_\mathrm{m0}=0.282^{+0.160}_{-0.059}$,  $w_0=-0.91^{+0.18}_{-0.14}$ and $w_a=-0.28^{+1.40}_{-0.63}$. 
		The symbol $\Delta$ denotes the differences between the results of $d_{L,i}$, $H_i$ and $q_i$, and the predictions of $\Lambda$CDM model (solid line) and the $w_0w_a$CDM model  (dashed line). 
		\label{fig1}
	}
\end{figure*}

In Fig.~\ref{fig1}, we show the constraints on $d_{L,i}$, $H_i$, and $q_i$ from the Pantheon+ sample.  
The gray solid lines represent  the evolutionary curves of $d_{L}(z)$, $H(z)$, and $q(z)$ in the $\Lambda$CDM model with $H_0=73.2\pm0.94~\mathrm{km~s^{-1}Mpc^{-1}}$ and $\Omega_\mathrm{m0}=0.33\pm0.018$, which are obtained from the Pantheon+ SN~Ia data. 
To show the difference between these parameters and  the predictions of the  $\Lambda$CDM model clearly, we also plot   $\Delta d_{L,i}\equiv d_{L,i}-d_{L,\mathrm{\Lambda CDM}}$, $\Delta H_i\equiv H_i-H_\mathrm{\Lambda CDM}$, and $\Delta q_i\equiv q_i-q_\mathrm{\Lambda CDM}$. 
The uncertainties of $\Delta d_{L,i}$, $\Delta H_i$, and $\Delta q_i$ are calculated using the error propagation formula, \textit{i.e.}, $\sigma_{\Delta d_{L,i}}^2=\sigma_{d_{L,i}}^2+\sigma_{d_{L,\mathrm{\Lambda CDM}}}^2$, $\sigma_{\Delta H_i}^2=\sigma_{H_i}^2+\sigma_{H_\mathrm{\Lambda CDM}}^2$, and $\sigma_{\Delta q_i}^2=\sigma_{q_i}^2+\sigma_{q_\mathrm{\Lambda CDM}}^2$.
Here, $\sigma_{d_{L,\mathrm{\Lambda CDM}}}$, $\sigma_{H_\mathrm{\Lambda CDM}}$, and $\sigma_{q_\mathrm{\Lambda CDM}}$  represent the uncertainties derived from the $\Lambda$CDM model. 
The corresponding numerical results are summarized in Table~\ref{tab2}. 
It is easy to see that all values of  $d_{L,i}$ are compatible with the $\Lambda$CDM model.
The values of $H_i$ align with the $\Lambda$CDM model at the first three redshift points (\textit{i.e.}, $z_i=0.12$, $0.22$, and $0.32$). However, they deviate from the model at the last two redshift points, with the largest deviation reaching about 1.8$\sigma$ CL at $z_i=0.42$.
For the deceleration parameter, only the value at $z_i=0.22$ slightly differs from the prediction of the model by about 1.3$\sigma$ CL.
We also plot the evolutionary curves of $d_{L}(z)$, $H(z)$, and $q(z)$ in the $w_0w_a$CDM model, shown as blue dashed lines in Fig.~\ref{fig1}.
For the $w_0w_a$CDM model,  the EoS of dark energy is given by the CPL parametrization ($w_\mathrm{DE}=w_0+w_a\frac{z}{1+z}$).
We set $H_0=72.98\pm0.94~\mathrm{km~s^{-1}Mpc^{-1}}$, $\Omega_\mathrm{m0}=0.282^{+0.160}_{-0.059}$,  $w_0=-0.91^{+0.18}_{-0.14}$ and $w_a=-0.28^{+1.40}_{-0.63}$, which are given by the Pantheon+ sample.
One can see that most values of $d_{L,i}$, $H_i$, and $q_i$ are consistent with the predictions of the $w_0w_a$CDM  model.  However, there are notable deviations:
 $d_{L,i}$ at $z_i=0.52$ and  $H_i$ at $z_i=0.42$ and $0.52$   differ from this model by more than 1$\sigma$ CL. 

\begin{figure}
	\centering{
		\includegraphics[width=0.95\columnwidth]{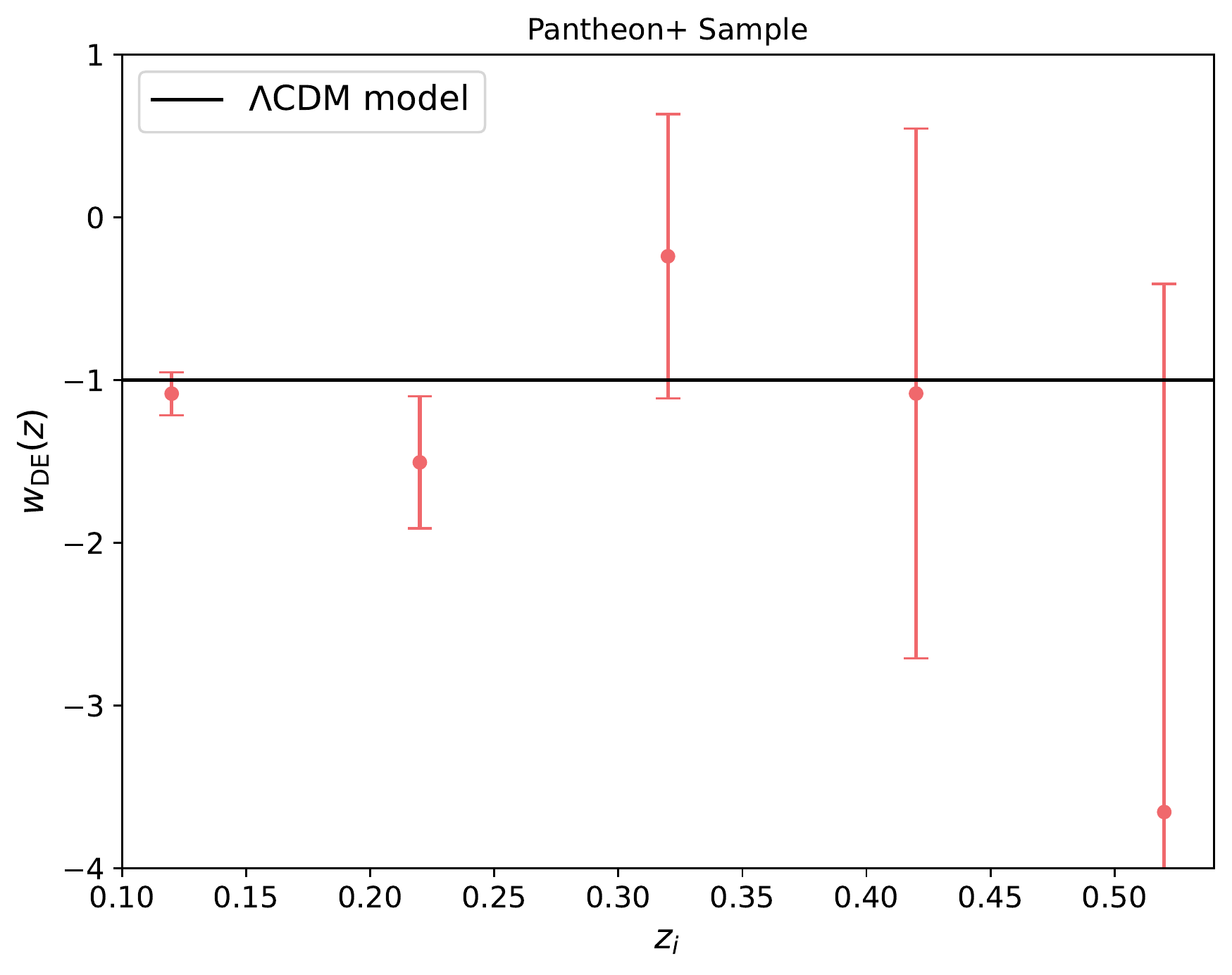}	}
	\caption{
		The values of $w_{\mathrm{DE},i}$ derived from Eq.~(\ref{eq:w}) with $H_0=73.2\pm0.94~\mathrm{km~s^{-1}Mpc^{-1}}$, $\Omega_\mathrm{m0}=0.33\pm0.018$ and the constraints on $H_i$ and $q_i$ from the Pantheon+ sample. 
		\label{fig2}
	}
\end{figure}

Using equation~(\ref{eq:w}), we can derive the EoS parameter of dark energy at different redshifts ($w_{\mathrm{DE},i}$) by considering the constraints on $H_i$ and $q_i$. 
Setting  $H_0=73.2\pm0.94~\mathrm{km~s^{-1}Mpc^{-1}}$ and $\Omega_\mathrm{m0}=0.33\pm0.018$, we obtain the values of $w_{\mathrm{DE},i}$ at five different redshifts, which are  shown in   Fig.~\ref{fig2}.
We find that  $w_{\mathrm{DE},i}$ at $z_i=0.22$  is slightly smaller than $-1$, and 
other values of  $w_{\mathrm{DE},i}$ agree well with  the $-1$ line within $1\sigma$ CL.  Figure~\ref{fig2} shows that $w_{\mathrm{DE},i}$ have very large uncertainties. This is because that only the SN Ia data in the redshift region $[z_i-\Delta z, z_i+\Delta z]$ are used and these  SN Ia  cannot provide tight constraints on $H_i$ and $q_i$, especially $q_i$, which result in the large uncertainties on $w_{\mathrm{DE},i}$. 

\begin{table*}
	\centering
	\caption{ 
		Number of data in each redshift range. The $N_\mathrm{Pantheon+}$, $N_\mathrm{P+1690}$ and $N_{H(z)}$ represent the number of Pantheon+ SN~Ia data, P+1690 SN~Ia data, and $H(z)$ data in each redshift range, respectively. }
	\label{tab1}
	\begin{threeparttable}
		\begin{tabular}{lccccc}
			\hline
			Redshift range & $z\leq0.24^\mathrm{a}$ & $0.1<z\leq0.34$ & $0.2<z\leq0.44$ & $0.3<z\leq0.54$ & $0.4<z\leq0.64$  \\
			\hline
			$N_\mathrm{Pantheon+}$ & 944 & 567 & 497 & 321 & 207   \\
			$N_\mathrm{P+1690}$ & 378 & 566 & 493 & 317 & 206   \\
			$N_{H(z)}$ & 7 & 7 & 7 & 9 & 7 \\
			\hline
		\end{tabular}
		\begin{tablenotes}
			\footnotesize
			\item[a] The data with $z<0.01$ are excluded for the Pantheon+ sample, and with $z<0.06$ are excluded for the P+1690 sample.
		\end{tablenotes}
	\end{threeparttable}
\end{table*}

\subsection{ P+1690 SN Ia}
It is noteworthy that the Pantheon+ team employed the SALT2mu algorithm within the SNANA package~\citep{Kessler2009}  to derive the corrected supernova sample, including the covariance matrix. This algorithm utilizes a fidial cosmological model, specifically selecting the $w$CDM model for this purpose. Furthermore, the velocity field reconstruction~\citep{Carrick2015,Peterson2022}, employed to adjust the redshift from $z_\mathrm{CMB}$ to $z_\mathrm{HD}$ within the Pantheon+ sample, also requires the use of a fiducial model. Although this dependency on the fiducial model has been observed to be weak~\citep{Carr2022}.
Recently, an analysis of the Pantheon+ supernova catalog led to the creation of a truly cosmological-model-independent SN Ia sample~\cite{Lane203}, comprising 1690 data points, referred to as the P+1690 sample in our study. Unlike the Pantheon+ sample, the P+1690 sample does not include bias corrections tied to any cosmological model, nor does its covariance matrix of light-curve parameters ($C_\mathrm{P+1690}$) reflect uncertainties dependent on a cosmological model. Additionally, the P+1690 sample uses $z_\mathrm{CMB}$ rather than $z_\mathrm{HD}$, thereby not relying on velocity field reconstruction.

To accommodate the P+1690 sample, which only provides the light-curve parameters  $\{ m_B,~x_1,~c \}$ with their covariance matrix $C^{\eta}_\mathrm{P+1690}$,  
we need to adapt our calculation of the corrected distance modulus $\bm{\hat{\mu}}_\mathrm{obs}^\mathrm{corr}$ . This adjustment involves reformulating the relevant equations to work with the data format provided by the P+1690 sample~\citep{Betoule2014}:
\begin{eqnarray}
	\bm{\hat{\mu}}_\mathrm{obs}^\mathrm{corr} = \hat{\bm{A}} \hat{\bm{\eta}} + \hat{\bm{\delta}}_\mathrm{host} - M_B,
\end{eqnarray}   
where the correction $\delta_\mathrm{bias}$ is omitted.
Here $\hat{\bm{A}} = \hat{\bm{A}}_0 +\alpha \hat{\bm{A}}_1 - \beta\hat{\bm{A}}_2 $  with $(\hat{\bm{A}}_k)_{i,j}\equiv\delta_{3i+k,j}$, and  $\hat{\bm{\eta}}=\{ (m_{B})_1,~(x_{1})_1,~(c)_{1},\cdots,~(m_{B})_N,~(x_{1})_N,~(c)_{N} \}$ is the one-dimension light-curve parameter vector of  $N$ data points and consists of $3\times N$ elements. 
The $\hat{\bm{\delta}}_\mathrm{host}$ is a vector consisting of the following elements~\citep{Peterson2022}: 
\begin{eqnarray}
	\delta_\mathrm{host} = 
	\begin{cases}
		-\Delta_\mathrm{host}, & \text{if } M_\star < 10^{10}M_\odot \\
		+\Delta_\mathrm{host},  & \text{if } M_\star \geq 10^{10}M_\odot.
	\end{cases}
\end{eqnarray}
Here $M_\star$ and $M_\odot$ are the host-galaxy mass and the solar mass, respectively, and $\Delta_\mathrm{host}$ is a constant.  
The covariance matrix of the corrected distance modulus can be derived from
\begin{eqnarray}
	C_\mathrm{P+1690}=\hat{\bm{A}} C^\eta_\mathrm{P+1690} \hat{\bm{A}}^\dagger.
\end{eqnarray}

Replacing $C_\mathrm{P}$ with $C_\mathrm{P+1690}$ in Eq.~(\ref{eq:chi}), we can obtain  constraints on the cosmological parameters ($d_{L,i}$, $H_i$, $q_i$) from the  P+1690 sample after setting $M_B=-19.253\pm0.027~\mathrm{mag}$~\citep{Riess2022}. 
In our analysis of the P+1690 sample, we consider the coefficients  ($\alpha$, $\beta$, $\Delta_\mathrm{host}$)  as free parameters. To robustly account for their potential effects on the derived cosmological parameters, we simultaneously constrain these coefficients \footnote{
	In the MCMC analysis, we set the prior distribution of the coefficients ($\alpha$, $\beta$, and $\Delta_\mathrm{host}$) to be uniform with ranges  $0\leq\alpha\leq0.5$, $0\leq\beta\leq10$, and $-1\leq\Delta_\mathrm{host}\leq1$, respectively, and 
 obtain $\alpha=0.217\pm0.008$, $\beta=5.40^{+0.15}_{-0.17}$, and $\Delta_\mathrm{host}=0.031\pm0.006$ for the $\Lambda$CDM model.
	}
 and treat them as nuisance parameters through the marginalization method. 
Since the peculiar velocities can impact the SN Ia up to $z_\mathrm{CMB}\approx 0.06$~\citep{Davis2011}, for the P+1690 sample,  data with $z_\mathrm{CMB}<0.06$ are excluded in our discussions.

\begin{figure*}
	\centering
	\includegraphics[width=0.66\columnwidth]{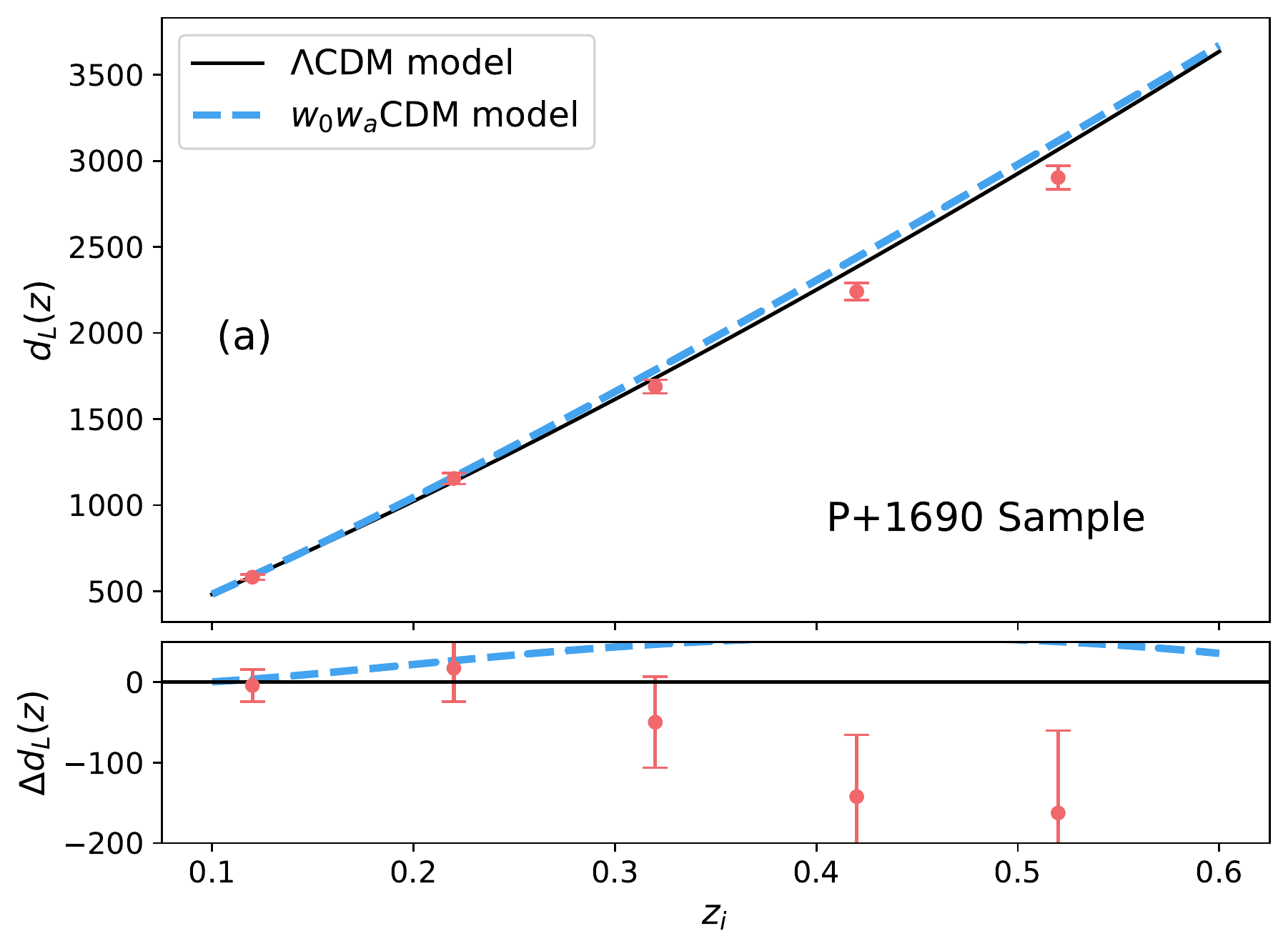}
	\includegraphics[width=0.66\columnwidth]{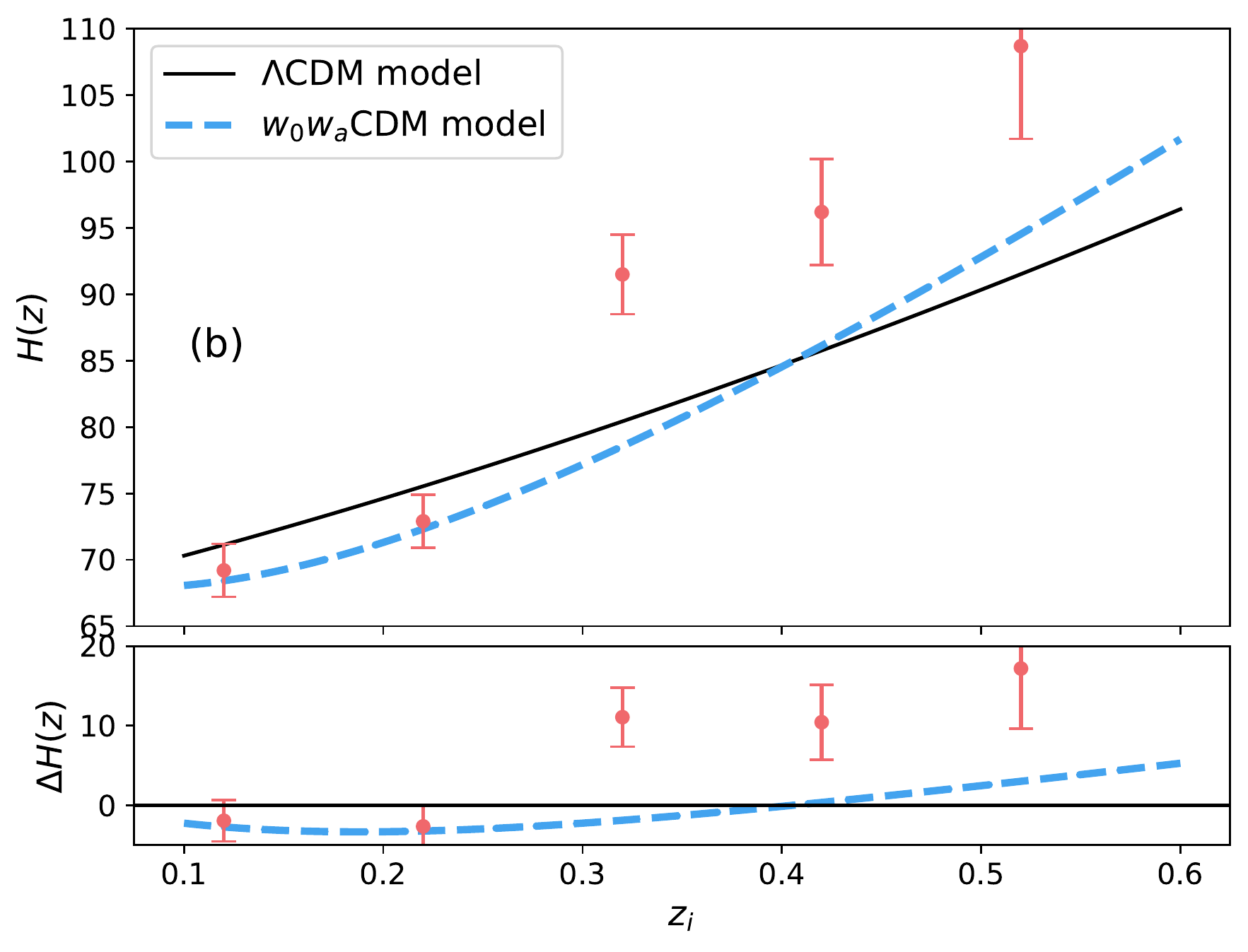}
	\includegraphics[width=0.66\columnwidth]{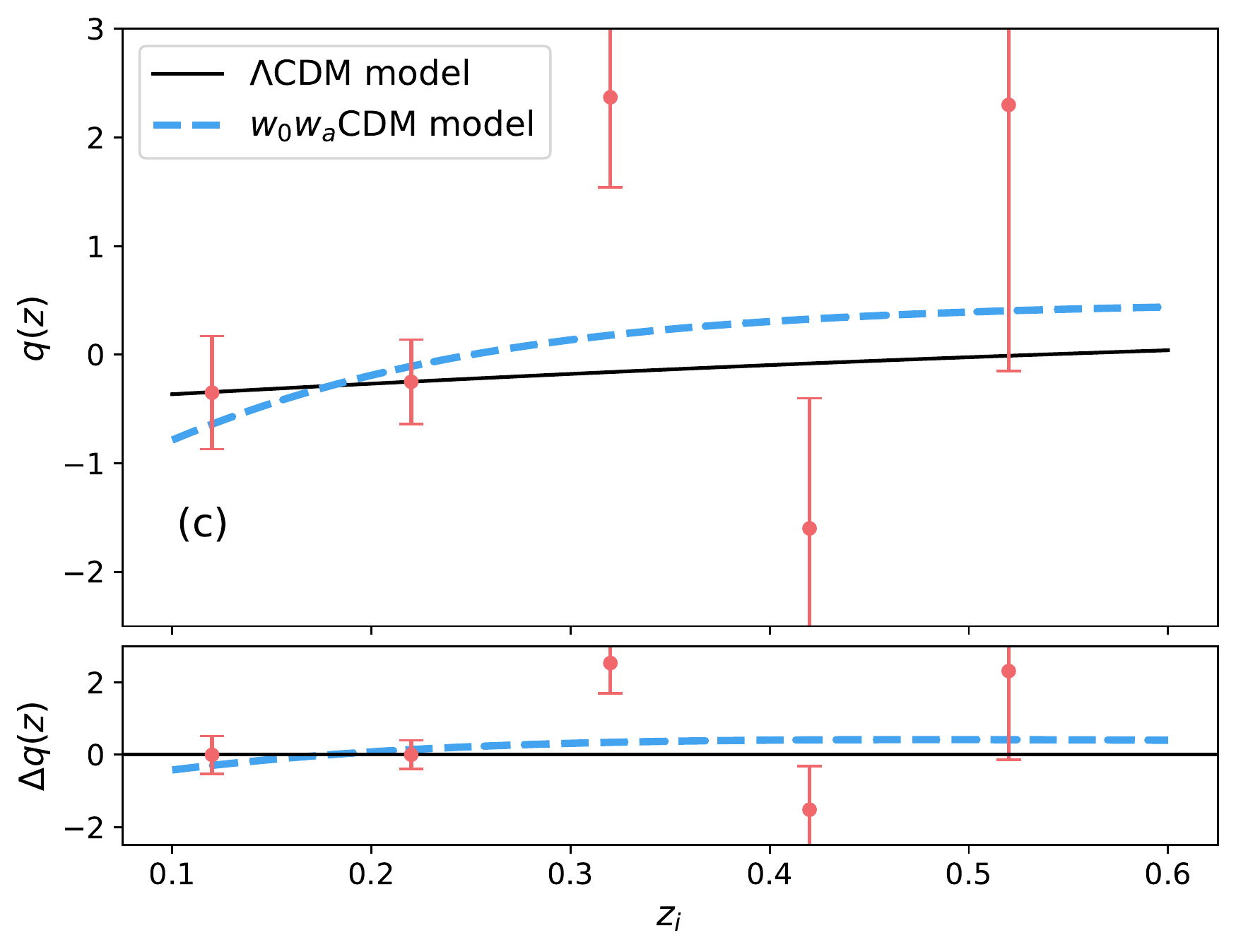}
	\caption{The constraints on $d_{L,i}$, $H_i$ and $q_i$ at different redshift points by using the P+1690 sample. 
	The black solid and blue dashed lines represent, respectively, the predictions of the $\Lambda$CDM model with $H_0=66.5\pm1.5~\mathrm{km~s^{-1}Mpc^{-1}}$ and $\Omega_\mathrm{m0}=0.356\pm0.032$, and the $w_0w_a$CDM model with $H_0=69.7\pm 2.1~\mathrm{km~s^{-1}Mpc^{-1}}$, $\Omega_\mathrm{m0}=0.515^{+0.045}_{-0.018}$, $w_0=-3.06^{+0.94}_{-0.75}$ and $w_a=0.5^{+4.3}_{-2.5}$. 
	The symbol $\Delta$ denotes the differences between the results of $d_{L,i}$, $H_i$ and $q_i$, and the predictions of the $\Lambda$CDM model (solid line) and the $w_0w_a$CDM model  (dashed line). 
		\label{fig3}
	}
\end{figure*}

In  Fig.~\ref{fig3}, we show constraints on the parameters $d_{L,i}$, $H_i$, and $q_i$ from the P+1690 sample. 
The black solid and blue dashed lines represent, respectively, the evolutions of $d_{L}(z)$, $H(z)$ and $q(z)$ in the $\Lambda$CDM model with $H_0=66.5\pm1.5~\mathrm{km~s^{-1}Mpc^{-1}}$ and $\Omega_\mathrm{m0}=0.356\pm0.032$, and the $w_0w_a$CDM model with $H_0=69.7\pm 2.1~\mathrm{km~s^{-1}Mpc^{-1}}$, $\Omega_\mathrm{m0}=0.515^{+0.045}_{-0.018}$, $w_0=-3.06^{+0.94}_{-0.75}$ and $w_a=0.5^{+4.3}_{-2.5}$, which are given by the P+1690 sample.
The corresponding numerical results are also summarized in the upper part of Table~\ref{tab3}.  It is easy to see that the allowed values of $d_{L,i}$ in the first three redshifts agree with the predictions of the $\Lambda$CDM model, while  deviations from this model  appear at more than 1$\sigma$ CL for the last two redshift points (\textit{i.e.}, $z_i=0.42$ and $0.52$). This result is different from that obtained from the Pantheon+ sample where it has been found that all values of $d_{L,i}$ are compatible with those derived from the  $\Lambda$CDM model.  Moreover, the consistency between $d_{L,i}$ and the $w_0w_a$CDM model deteriorates when utilizing the P+1690 sample. For $H_i$, deviations from both the $\Lambda$CDM and $w_0w_a$CDM models are observed for the last three redshift points, contrary to the results shown in Fig.~\ref{fig1}(b), with the greatest deviation at $z_i=0.32$ reaching approximately 3$\sigma$ CL.
A deviation with  similar statistical significance can also be observed in Fig.~\ref{fig3}(c) where the  constraints on $q_i$ are plotted, and this deviation occurs  at $z_i=0.32$ too, which  differs from those obtained from the Pantheon+ sample where the results are consistent with the $\Lambda$CDM model and the $w_0w_a$CDM model within 2$\sigma$ CL. 

\begin{figure}
	\centering{
		\includegraphics[width=\columnwidth]{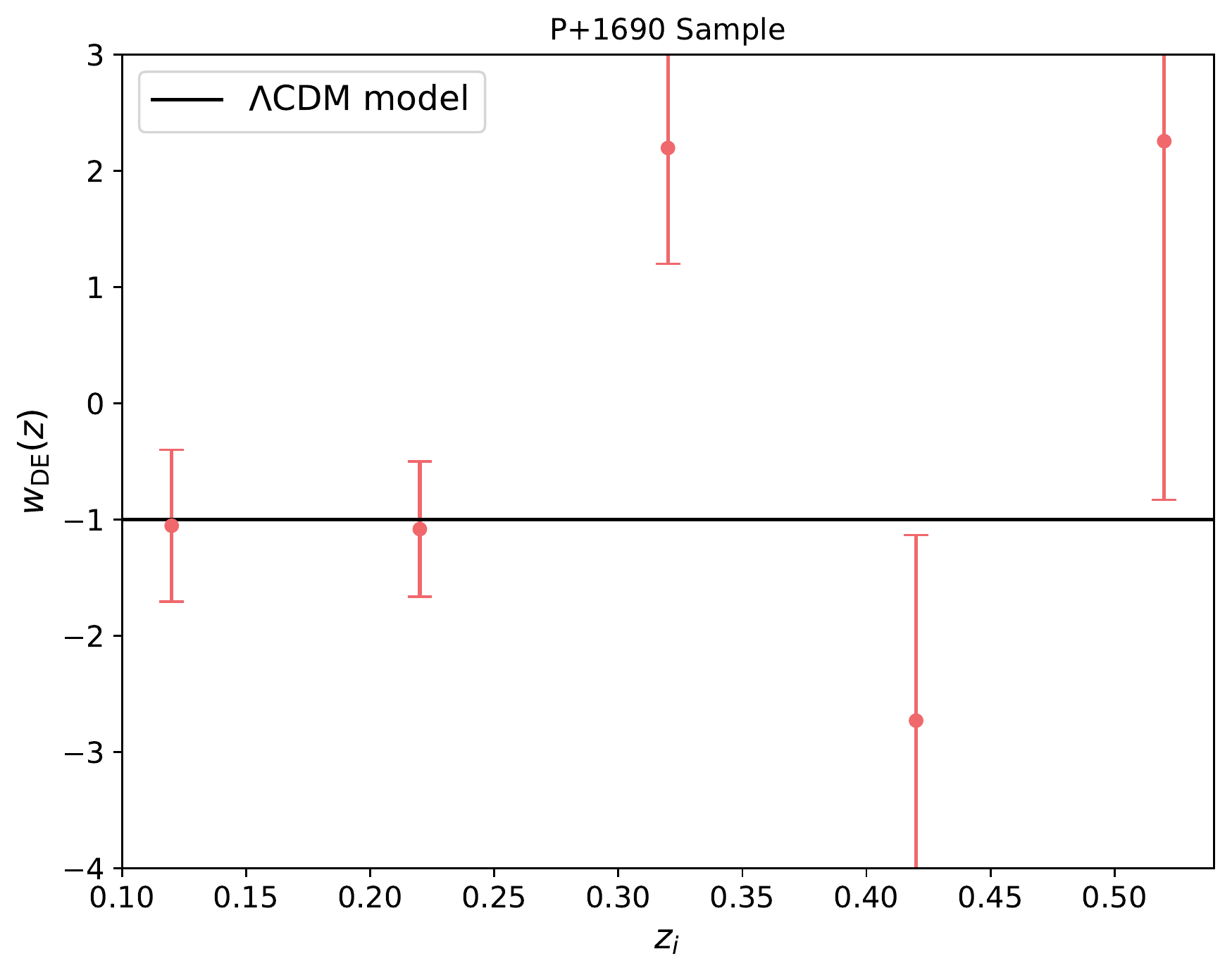}
	}
	\caption{
		The values of $w_{\mathrm{DE},i}$ derived from Eq.~(\ref{eq:w}) with $H_0=66.5\pm1.5~\mathrm{km~s^{-1}Mpc^{-1}}$,  $\Omega_\mathrm{m0}=0.356\pm0.032$ and the constraints on $H_i$ and $q_i$ from the P+1690 sample. 
		\label{fig4}
	}
\end{figure}

Setting $H_0$ and $\Omega_\mathrm{m0}$ to be $66.5\pm1.5~\mathrm{km~s^{-1}Mpc^{-1}}$ and $0.356\pm0.032$, we  derive the values of $w_{\mathrm{DE},i}$ at five different redshifts, as shown in Fig.~\ref{fig4}. 
We find that except for $z_i=0.32$, all $w_{\mathrm{DE},i}$ align within a 2$\sigma$ CL of the expected $w_\mathrm{DE}=-1$ value. 
The value of $w_{\mathrm{DE},i}$ at $z_i=0.32$ deviates from  $-1$  by about 3$\sigma$ CL, which is different from what is obtained from the Pantheon+ sample. 
Here, we must emphasize that the uncertainties of $w_{\mathrm{DE},i}$  are quite high due to the lack of constraining data.

\begin{table}
	\centering
	\caption{ Constraints on parameters in each redshift range using the Pantheon+ sample with and without the $H(z)$ data. 
		The marginalized mean values with 1$\sigma$ uncertainty of  parameters are shown.
		$\Delta$ denotes the difference between the constraint results and the  $\Lambda$CDM model. }
	\label{tab2}
	\scriptsize
	\begin{tabular}{ccccc}
		\hline
		\hline
		\multicolumn{5}{c}{\textbf{Pantheon+~Sample}}\\
		\hline
		$z_i$ & $d_{L,i}$ & $H_i$  & $q_i$ & $w_{\mathrm{DE},i}$ \\
		\hline
		$0.12$ & $531\pm7$ & $78.5\pm1.1$ & $-0.47\pm0.11$ & $-1.08\pm0.13$ \\
		$0.22$ & $1034\pm13$ & $82.0\pm1.3$ & $-0.68\pm0.31$ & $-1.51\pm0.41$ \\
		$0.32$ & $1593\pm21$ & $87.9\pm2.0$ & $0.33\pm0.62$ & $-0.24\pm0.87$ \\
		$0.42$ & $2167\pm30$ & $99.9\pm3.6$ & $-0.3\pm1.20$ & $-1.08\pm1.63$ \\
		$0.52$ & $2770\pm40$ & $105.1\pm5.4$ & $-1.90\pm2.10$ & $-3.65\pm3.25$ \\
		\hline
		$z_i$ & $\Delta d_{L,i}$ & $\Delta H_i$ & $\Delta q_i$ & $\Delta w_{\mathrm{DE},i}$ \\
		\hline
		$0.12$ & $-3\pm10$ & $0.6\pm1.5$ & $-0.08\pm0.11$  & $-0.08\pm0.13$ \\
		$0.22$ & $-4\pm19$ & $-0.5\pm1.7$ & $-0.39\pm0.31$ & $-0.50\pm0.41$ \\
		$0.32$ & $4\pm30$ & $0.4\pm2.4$ & $0.53\pm0.62$ & $0.76\pm0.87$ \\
		$0.42$ & $-14\pm43$ & $6.9\pm3.9$ & $-0.18\pm1.20$ & $-0.08\pm1.63$ \\
		$0.52$ & $-40\pm57$ & $6.1\pm5.7$ & $-1.85\pm2.10$ & $-2.65\pm3.25$ \\
		\hline
		\hline
		\multicolumn{5}{c}{\textbf{Pantheon+~Sample Plus $H(z)$ Data}}\\
		\hline
		$z_i$ & $d_{L,i}$ & $H_i$ & $q_i$ & $w_{\mathrm{DE},i}$ \\
		\hline
		$0.12$ & $568\pm20.5$ & $73.4\pm2.6$ & $-0.47\pm0.11$ & $-1.06\pm0.14$ \\
		$0.22$ & $1108\pm42$ & $76.6\pm2.9$ & $-0.69\pm0.31$ & $-1.48\pm0.41$ \\
		$0.32$ & $1801\pm130$ & $78.1\pm5.1$ & $0.29\pm0.55$ & $-0.32\pm0.85$ \\
		$0.42$ & $2586\pm155$ & $83.7\pm4.4$ & $-0.4\pm1.0$ & $-1.56\pm1.76$ \\
		$0.52$ & $3354\pm250$ & $89.3\pm6.5$ & $0.0\pm1.1$ & $-0.99\pm2.19$ \\
		\hline
		$z_i$ & $\Delta d_{L,i}$ & $\Delta H_i$ & $\Delta q_i$ & $\Delta w_{\mathrm{DE},i}$ \\
		\hline
		$0.12$ & $-10.9\pm25.2$ & $1.5\pm3.2$ & $-0.08\pm0.11$ & $-0.06\pm0.14$ \\
		$0.22$ & $-18\pm51$ & $0.6\pm3.5$ & $-0.40\pm0.31$ & $-0.48\pm0.41$ \\
		$0.32$ & $78\pm137$ & $-2.6\pm5.5$ & $0.49\pm0.55$ & $0.68\pm0.85$ \\
		$0.42$ & $220\pm166$ & $-2.1\pm5.0$ & $-0.28\pm1.00$ & $-0.56\pm1.76$ \\
		$0.52$ & $307\pm262$ & $-2.0\pm7.0$ & $0.05\pm1.10$ & $0.01\pm2.19$ \\
		\hline
	\end{tabular}
	
\end{table}

\begin{table}
	\centering
	\caption{ Constraints on parameters in each redshift range using the P+1690 sample with and without the $H(z)$ data.
	The marginalized mean values with 1$\sigma$  uncertainty of parameters are shown. 
	$\Delta$ denotes the difference between the constraint results and the  $\Lambda$CDM model. }
	\label{tab3}
	\scriptsize
	\begin{tabular}{ccccc}
		\hline
		\hline
		\multicolumn{5}{c}{\textbf{P+1690~Sample}}\\
		\hline
		$z_i$ & $d_{L,i}$ & $H_i$  & $q_i$ & $w_{\mathrm{DE},i}$ \\
		\hline
		$0.12$ & $582\pm15$ & $69.2\pm2.0$ & $-0.35\pm0.52$ & $-1.05\pm0.65$ \\
		$0.22$ & $1155\pm32$ & $72.9\pm2.0$ & $-0.25\pm0.39$ & $-1.08\pm0.58$ \\
		$0.32$ & $1689\pm39$ & $91.5\pm3.0$ & $2.37\pm0.83$ & $2.20\pm1.00$ \\
		$0.42$ & $2241\pm50$ & $96.2\pm4.0$ & $-1.60\pm1.20$ & $-2.73\pm1.60$ \\
		$0.52$ & $2903\pm68$ & $108.7\pm7.0$ & $2.30\pm2.45$ & $2.26\pm3.09$ \\ 
		\hline
		$z_i$ & $\Delta d_{L,i}$ & $\Delta H_i$ & $\Delta q_i$ & $\Delta w_{\mathrm{DE},i}$ \\
		\hline
		$0.12$ & $-4\pm20$ & $-1.9\pm2.6$ & $-0.005\pm0.523$ & $-0.05\pm0.65$  \\
		$0.22$ & $17\pm41$ & $-2.6\pm2.7$ & $-0.001\pm0.393$ & $-0.08\pm0.58$ \\ 
		$0.32$ & $-50\pm57$ & $11.1\pm3.7$ & $2.53\pm0.83$ & $3.20\pm1.00$ \\
		$0.42$ & $-142\pm77$ & $10.4\pm4.7$ & $-1.52\pm1.20$ & $-1.73\pm1.60$ \\
		$0.52$ & $-163\pm102$ & $17.2\pm7.6$ & $2.31\pm2.45$ & $3.26\pm3.09$ \\
		\hline
		\hline
		\multicolumn{5}{c}{\textbf{P+1690~Sample Plus $H(z)$ Data}}\\
		\hline
		$z_i$ & $d_{L,i}$ & $H_i$ & $q_i$ & $w_{\mathrm{DE},i}$ \\
		\hline
		$0.12$ & $552\pm25$ & $73.0\pm3.2$ & $-0.37\pm 0.49$ & $-0.98\pm0.56$ \\
		$0.22$ & $1089\pm41$ & $77.3\pm 2.9$ & $-0.26\pm 0.38$ & $-0.96\pm0.49$ \\
		$0.32$ & $2098\pm 170$ & $73.2\pm5.2$ & $1.94\pm 0.77$& $2.90\pm1.86$ \\
		$0.42$ & $2582\pm\pm165$ & $83.9\pm 4.5$ & $-1.40\pm 1.00$ & $-3.46\pm2.03$ \\
		$0.52$ & $3319\pm250$ & $93.2\pm 6.4$ & $1.04\pm1.00$ & $0.97\pm1.82$ \\
		\hline
		$z_i$ & $\Delta d_{L,i}$ & $\Delta H_i$ & $\Delta q_i$ & $\Delta w_{\mathrm{DE},i}$ \\
		\hline
		$0.12$ & $-30\pm30$ & $1.4\pm3.9$ & $-0.01\pm0.49$ & $0.02\pm0.56$ \\
		$0.22$ & $-42\pm53$ & $1.4\pm3.8$ & $0.003\pm0.383$ & $0.04\pm0.49$ \\
		$0.32$ & $369\pm178$ & $-7.5\pm5.8$ & $2.12\pm0.77$ & $3.90\pm1.86$ \\
		$0.42$ & $210\pm181$ & $-2.1\pm5.4$ & $-1.31\pm1.00$ & $-2.46\pm2.03$ \\
		$0.52$ & $267\pm268$ & $1.54\pm7.20$ & $1.06\pm1.00$ & $1.97\pm1.82$ \\
		\hline
	\end{tabular}

\end{table}

\subsection{SN Ia plus $H(z)$ data}
A prior fixed $M_B$ may introduce some unknown bias  in the  results.
To avoid this issue, we add  the Hubble parameter measurements into our analysis and then $M_B$ can be treated as a free parameter.
The latest $H(z)$ data determined from the cosmic chronometric technique \citep{Loeb1998,Jimenez2002} comprises 32 data points, covering redshifts ranging from 0.07 to 1.965~\citep{Simon2005,Stern2010,Moresco2012,Cong2014,Moresco2015,Moresco2016,Ratsimbazafy2017,Borghi2022, Wu2007}.
Here, we utilize only 19 data points that fall within the redshift range $z\leq 0.64$ (see Table \ref{tab1} for details).
To use the $H(z)$ data, we need to perform the Taylor expansion of the Hubble parameter similar to the equation (\ref{eq:dl}):
\begin{eqnarray}
	H(z) = H_i \bigg(1+(z-z_i) \frac{1+q_i}{1+z_i} \bigg) + \mathcal{O}\left((z-z_i)^2\right) \ . \label{eq:Hz}
\end{eqnarray}

\begin{figure*}
	\centering
	\includegraphics[width=0.66\columnwidth]{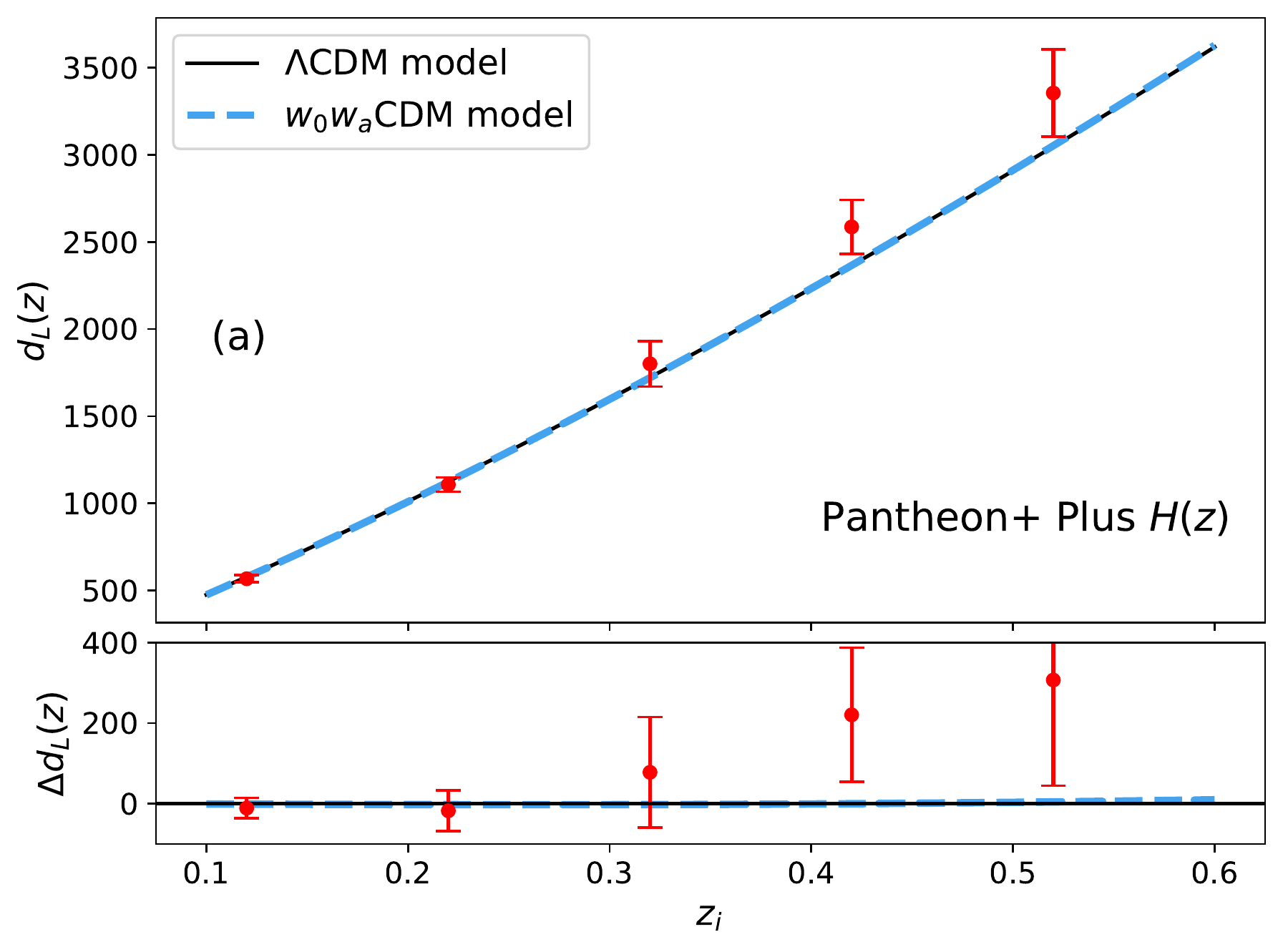}
	\includegraphics[width=0.66\columnwidth]{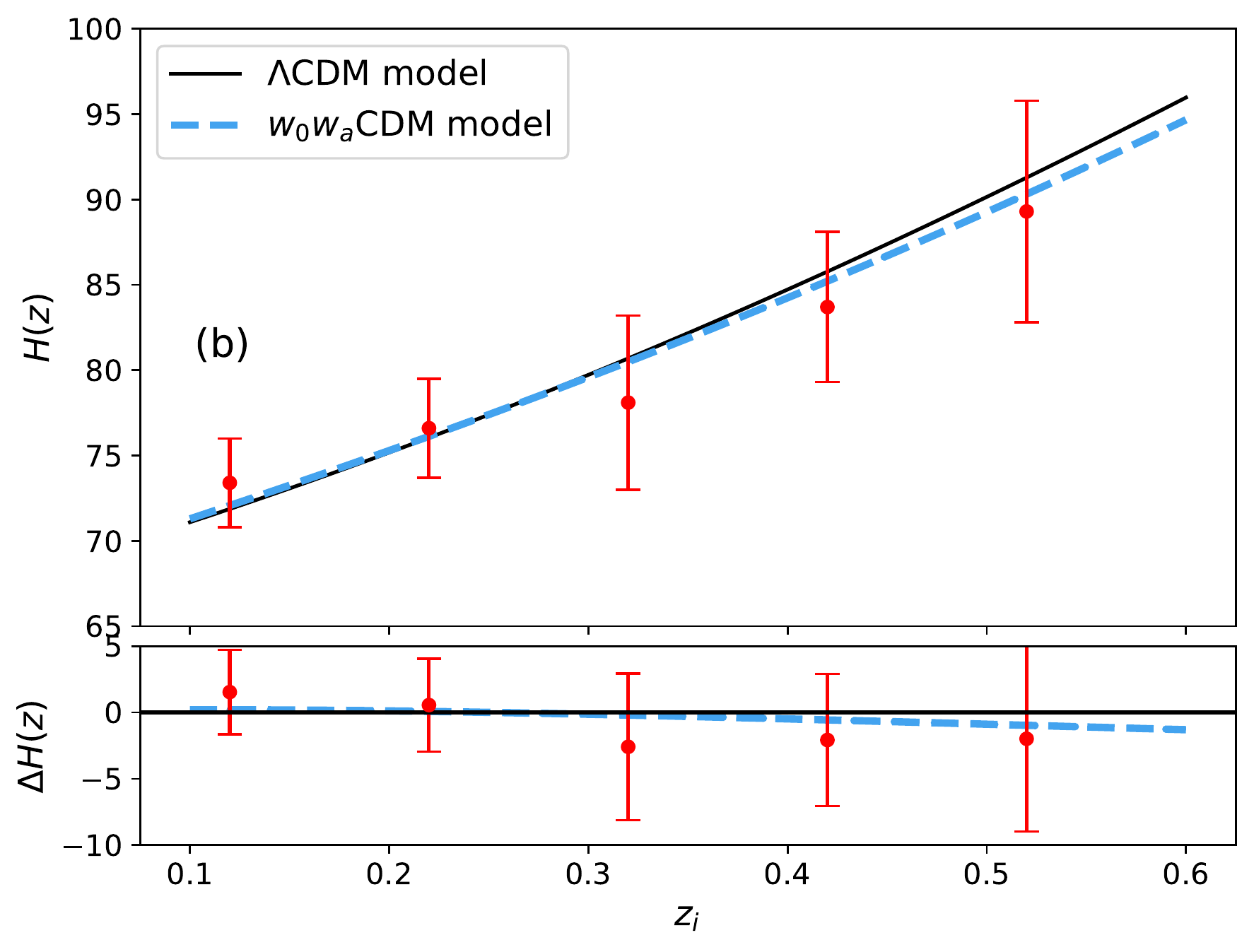}
	\includegraphics[width=0.66\columnwidth]{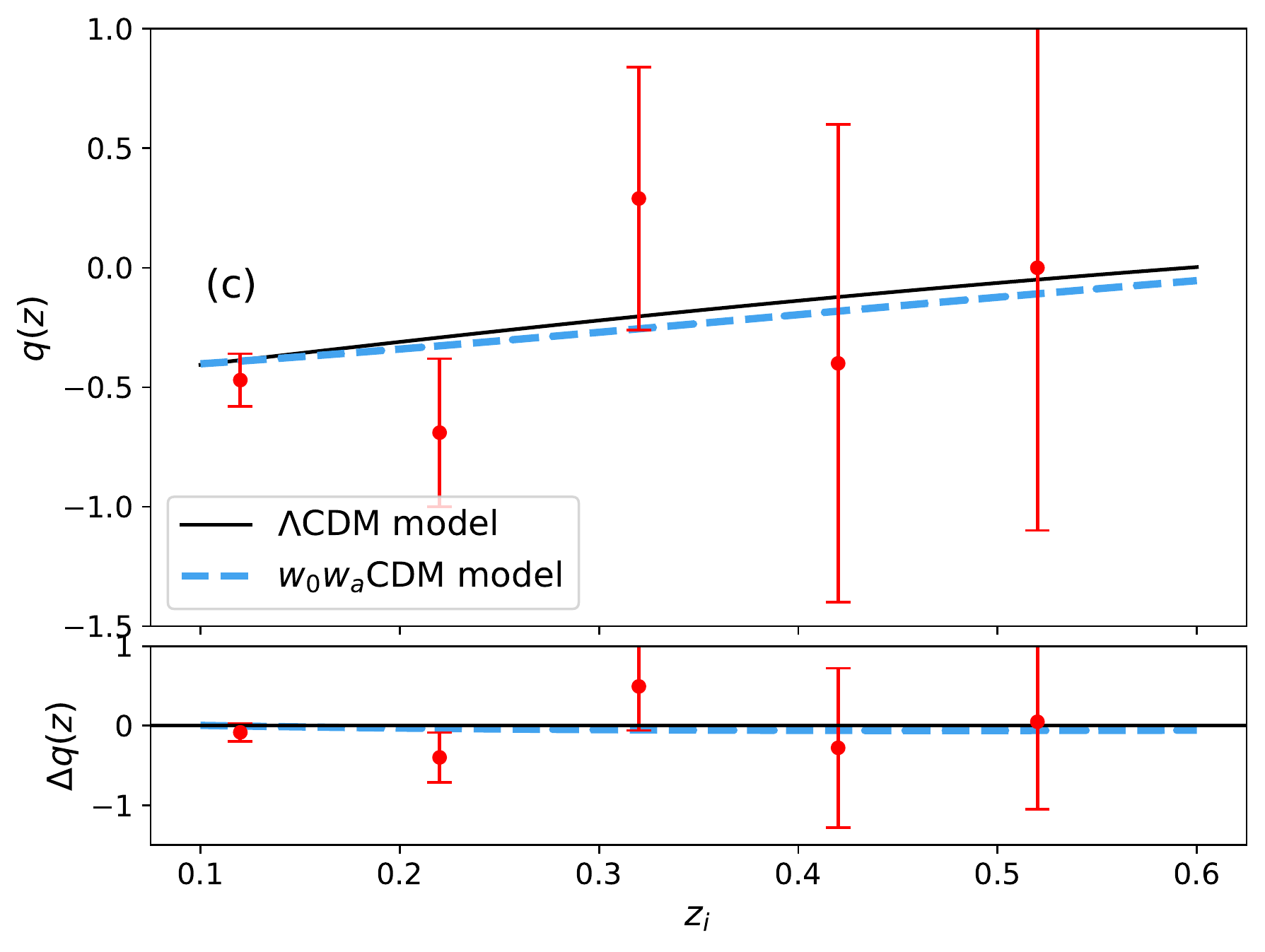}
	\caption{
		The constraints on $d_{L,i}$, $H_i$ and $q_i$ at different redshift points by using the Pantheon+ sample plus $H(z)$ data. The black solid and blue dashed lines represent, respectively, the predictions of the $\Lambda$CDM model with $H_0=67.5\pm1.7~\mathrm{km~s^{-1}Mpc^{-1}}$ and $\Omega_\mathrm{m0}=0.33\pm0.017$, and the $w_0w_a$CDM model with $H_0=67.5\pm 1.8~\mathrm{km~s^{-1}Mpc^{-1}}$, $\Omega_\mathrm{m0}=0.320^{+0.095}_{-0.045}$, $w_0=-0.93^{+0.12}_{-0.10}$ and $w_a=-0.47^{+1.4}_{-0.71}$. The symbol $\Delta$ denotes the differences between the results of $d_{L,i}$, $H_i$ and $q_i$, and the predictions of the $\Lambda$CDM model (solid line) and the $w_0w_a$CDM model  (dashed line). 
		\label{fig5}
	}
\end{figure*}

The results from Pantheon+ plus $H(z)$ data are shown in Fig.~\ref{fig5} and the lower part of Table \ref{tab2}.   
This figure  presents  $d_{L,i}$ $H_i$, $q_i$, and the predicted $d_L(z)$, $H(z)$, $q(z)$ from the $\Lambda$CDM model with $H_0=67.5\pm1.7~\mathrm{km~s^{-1}Mpc^{-1}}$ and $\Omega_\mathrm{m0}=0.33\pm0.017$, and the $w_0w_a$CDM model with $H_0=67.5\pm 1.8~\mathrm{km~s^{-1}Mpc^{-1}}$, $\Omega_\mathrm{m0}=0.320^{+0.095}_{-0.045}$, $w_0=-0.93^{+0.12}_{-0.10}$ and $w_a=-0.47^{+1.4}_{-0.71}$, respectively. 
From Fig.~\ref{fig5}(a), we find that, different from the results obtained from the Pantheon+ only, the values of $d_{L,i}$ at $z_i=0.42$ and $0.52$  deviate from the $\Lambda$CDM model. Figure~\ref{fig5}(b) indicates  that  the values of $H_i$ are consistent with those from the $\Lambda$CDM model at 1$\sigma$ CL, which are also different from the results obtained from Pantheon+ only.
Similar to what are shown in Fig.~\ref{fig1}, the value of $q_i$ at $z_i=0.22$  deviates from that from the $\Lambda$CDM  model by about 1.3$\sigma$ CL. However, the uncertainties of $q_i$  are reduced significantly  when the Hubble parameter measurements are included. For example, the 1$\sigma$ uncertainty of $q_i$ at $z_i=0.52$ is reduced from $2.1$ to $1.1$ after adding the $H(z)$ data.
This is expected since the parameter $q_i$ appears in the first-order coefficient in the expansion of the Hubble parameter (Eq.~(\ref{eq:Hz})), making it more sensitive to $q_i$ than Eq.~(\ref{eq:dl}). 
Thus, the SN~Ia sample plus $H(z)$ data can provide tighter constraints on $q_i$ than the SN~Ia sample alone.

\begin{figure}
	\centering{
		\includegraphics[width=\columnwidth]{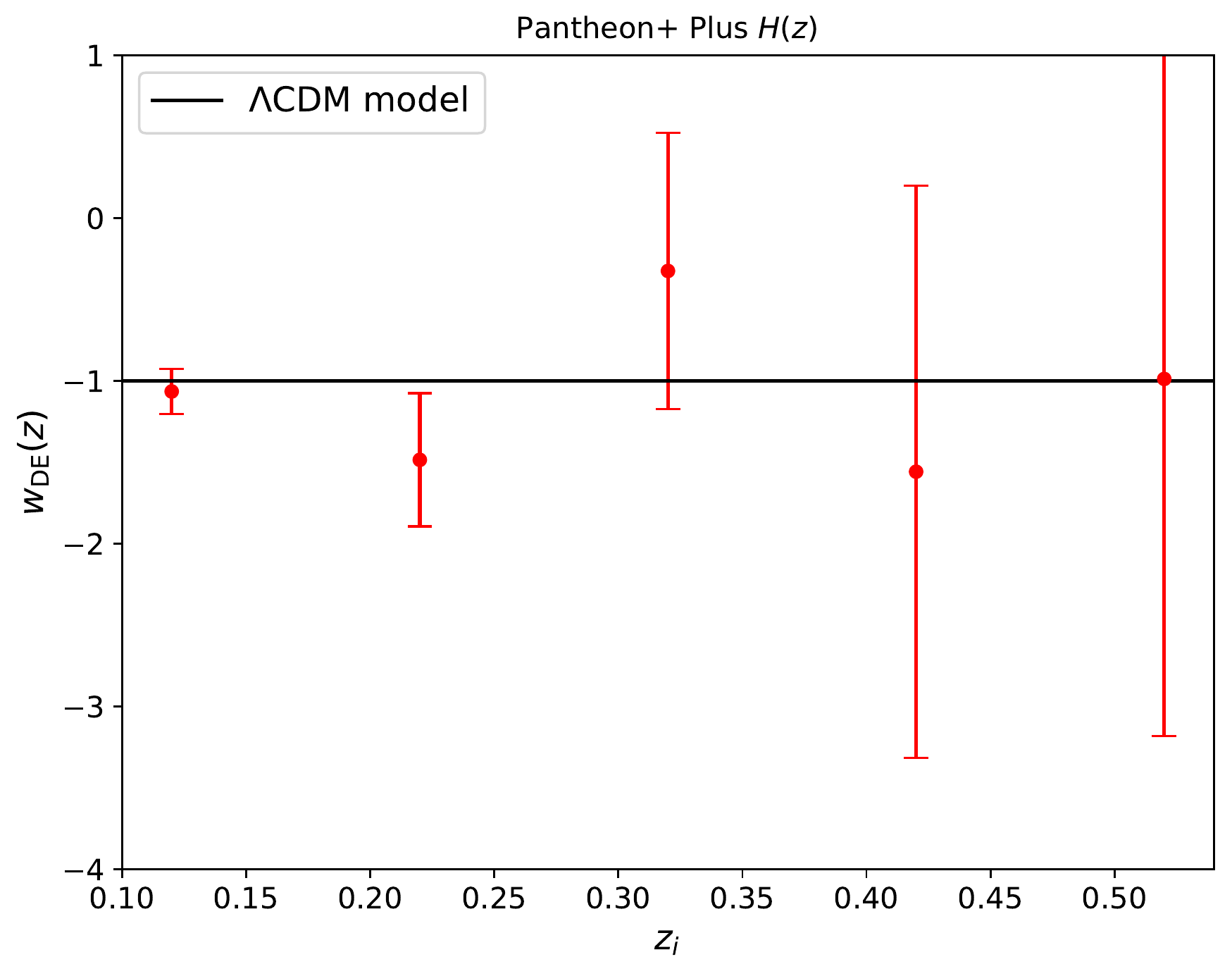}
	}
	\caption{
		The values of $w_{\mathrm{DE},i}$ derived from Eq.~(\ref{eq:w}) with $H_0=67.5\pm1.7~\mathrm{km~s^{-1}Mpc^{-1}}$, $\Omega_\mathrm{m0}=0.33\pm0.017$ and the constraints on $H_i$ and $q_i$ from the Pantheon+ sample plus $H(z)$ data. 
		\label{fig6}
	}
\end{figure}

Figure~\ref{fig6} shows the values of $w_{\mathrm{DE},i}$  after setting $H_0=67.5\pm1.7~\mathrm{km~s^{-1}Mpc^{-1}}$ and $\Omega_\mathrm{m0}=0.33\pm0.017$.
One can see that except for  the $w_{\mathrm{DE},i}$ at $z_i=0.22$ all other $w_{\mathrm{DE},i}$ align with the  $-1$ line within 1$\sigma$ CL. This is similar to what are shown in Fig.~\ref{fig2}.

\begin{figure*}
	\centering
	\includegraphics[width=.66\columnwidth]{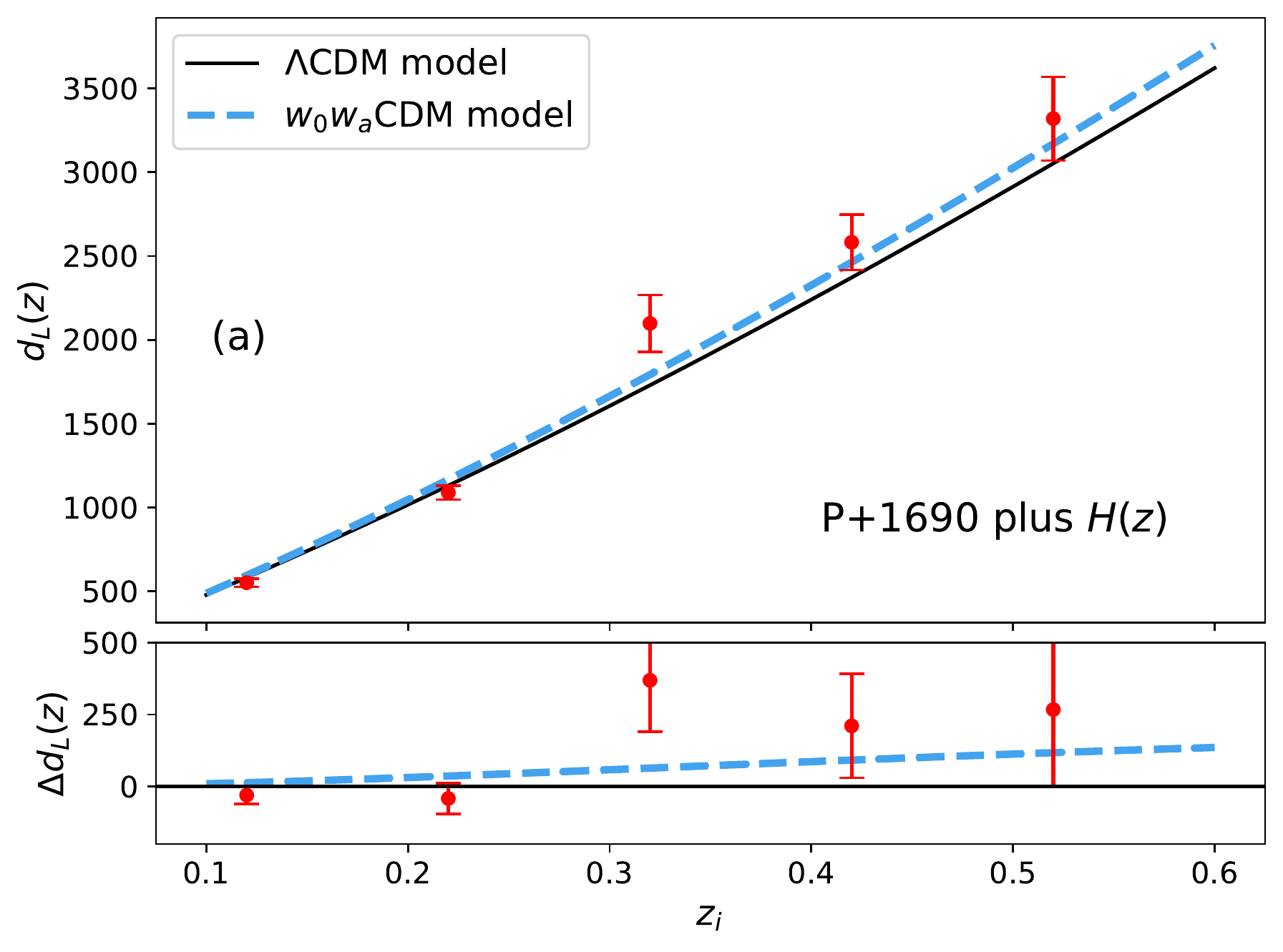}
	\includegraphics[width=.66\columnwidth]{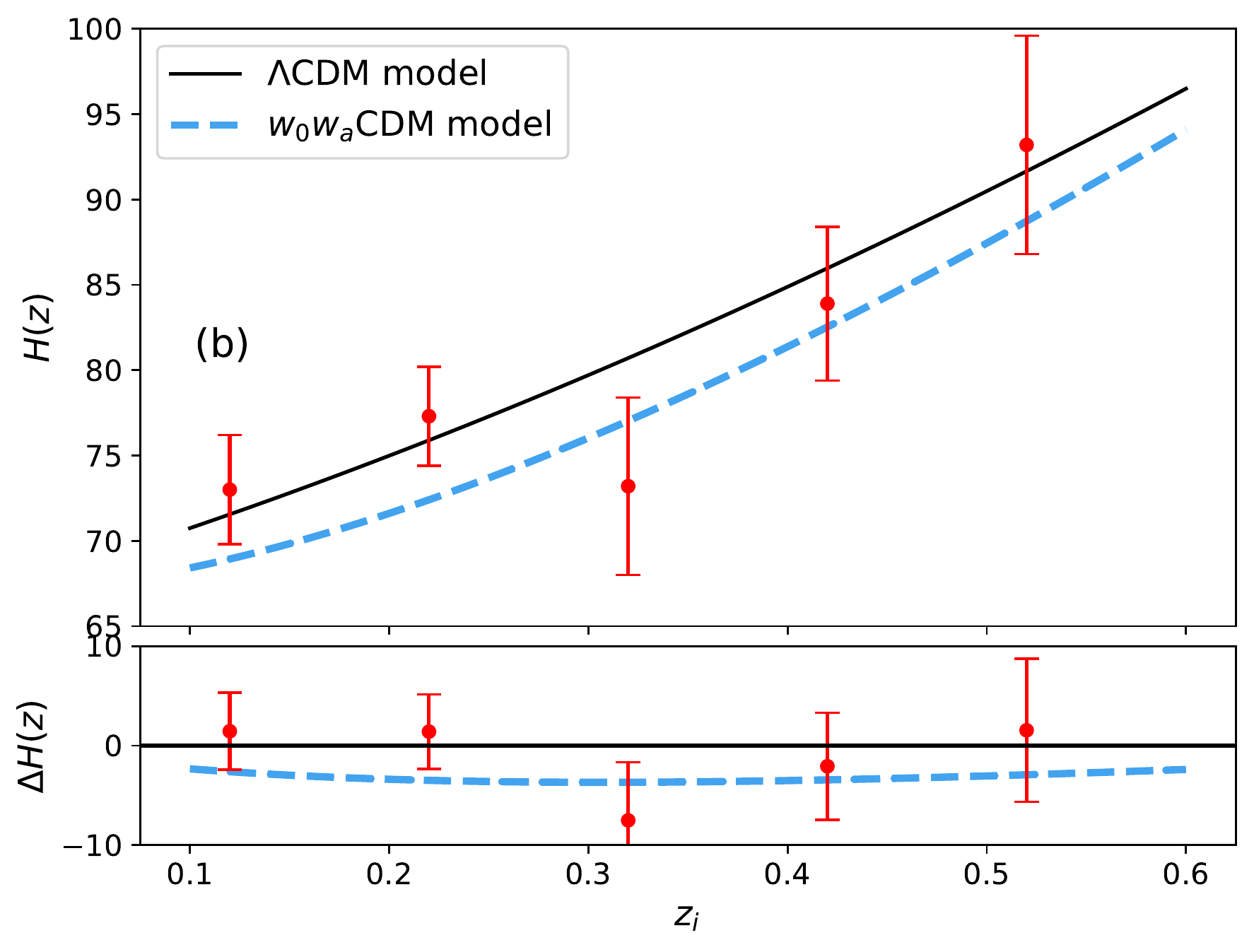}
	\includegraphics[width=.66\columnwidth]{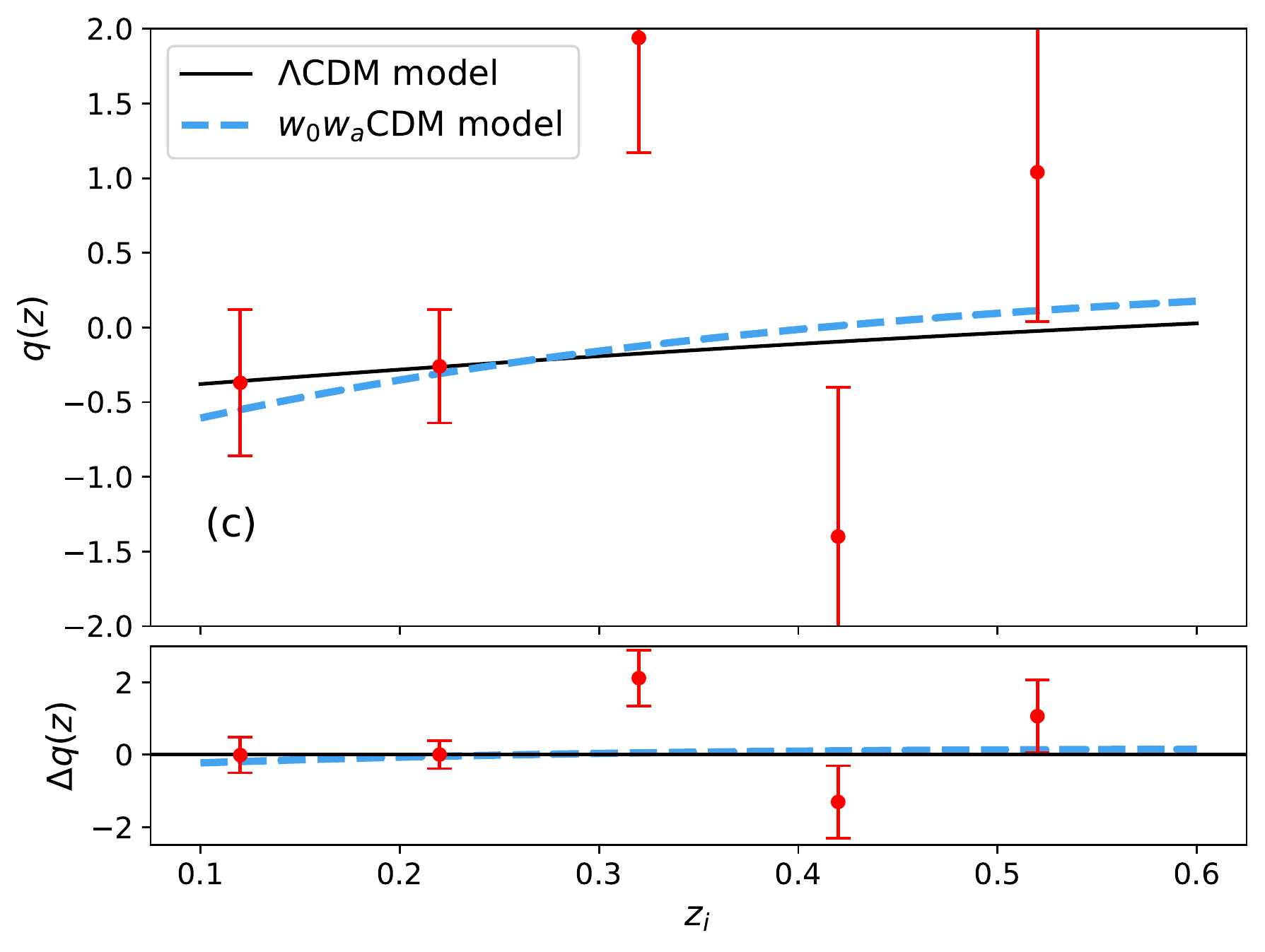}
	\caption{
		The constraints on $d_{L,i}$, $H_i$ and $q_i$ at different redshift points by using the P+1690 sample plus $H(z)$ data. 
		The black solid and blue dashed lines represent, respectively, the predictions of the $\Lambda$CDM model with $H_0=67.0\pm2.0~\mathrm{km~s^{-1}Mpc^{-1}}$ and $\Omega_\mathrm{m0}=0.347\pm0.03$, and the $w_0w_a$CDM model with $H_0=66.9\pm 2.0~\mathrm{km~s^{-1}Mpc^{-1}}$, $\Omega_\mathrm{m0}=0.396^{+0.092}_{-0.048}$, $w_0=-1.58^{+0.45}_{-0.26}$ and $w_a=1.02^{+1.6}_{-0.59}$. The symbol $\Delta$ denotes the differences between the results of $d_{L,i}$, $H_i$ and $q_i$, and the predictions of $\Lambda$CDM model  (solid line) and the $w_0w_a$CDM model  (dashed line). 
		\label{fig7}
	}
\end{figure*}

Figure~\ref{fig7} and the lower part of Table~\ref{tab3}  present constraints on the parameters obtained from the P+1690 sample plus $H(z)$ data. 
Their predicted values from the $\Lambda$CDM model with $H_0=67.0\pm2.0~\mathrm{km~s^{-1}Mpc^{-1}}$ and $\Omega_\mathrm{m0}=0.347\pm0.03$, and from the $w_0w_a$CDM model with $H_0=66.9\pm 2.0~\mathrm{km~s^{-1}Mpc^{-1}}$, $\Omega_\mathrm{m0}=0.396^{+0.092}_{-0.048}$, $w_0=-1.58^{+0.45}_{-0.26}$ and $w_a=1.02^{+1.6}_{-0.59}$ are plotted as black solid and blue dashed lines, respectively.
Contrary to the results obtained solely from P+1690 sample, the value of $d_{L,i}$ at $z_i=0.32$ deviates from the predictions  of the $\Lambda$CDM model by  approximately  2$\sigma$ CL.  However, at $z_i=0.52$, the value of $d_{L,i}$ aligns with the $\Lambda$CDM model, indicating consistency. 
Figure.~\ref{fig7}(b) indicates that   $H_i$ deviates from the predictions of the $\Lambda$CDM model only  at $z_i=0.32$.
The constraints on deceleration parameter $q_i$ are tighter than those obtained solely from P+1690 sample, but the value of $q_i$  at $z_i=0.32$ still deviates from that of the $\Lambda$CDM model by about 2.7$\sigma$ CL.
These results differ from those obtained from the Pantheon+ plus $H(z)$ data, in which all constraints on $d_{L,i}$, $H_i$, and $q_i$ are compatible with those from the $\Lambda$CDM model within 2$\sigma$ CL.
Furthermore, most of the constraints on $d_{L,i}$, $H_i$, and $q_i$ are compatible with the predictions of the $w_0w_a$CDM model  at 2$\sigma$ CL. But  $q_i$ at $z_i=0.32$ still shows a deviation of more than 2$\sigma$ CL.

\begin{figure}
	\centering{
		\includegraphics[width=\columnwidth]{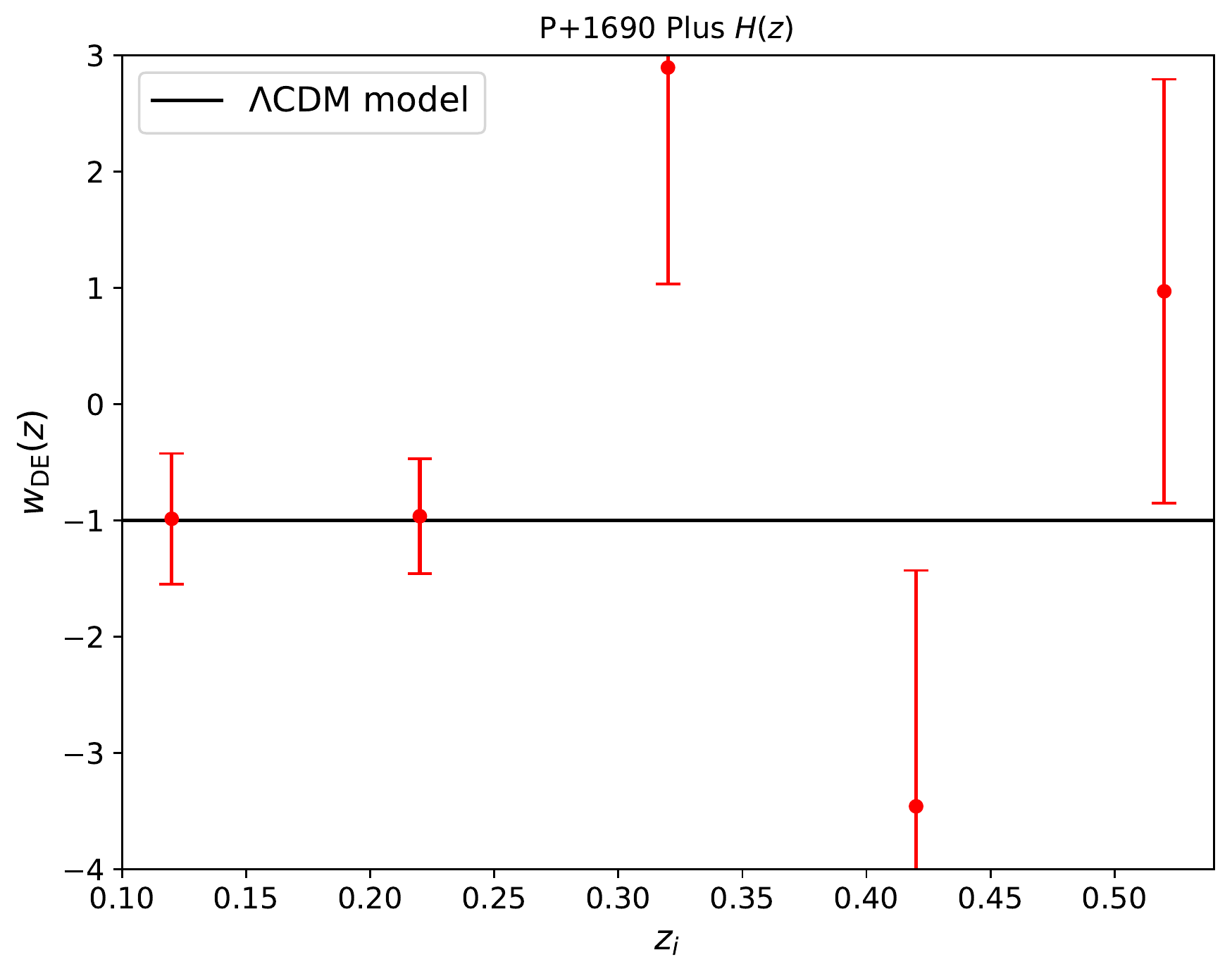}
	}
	\caption{
		The values of $w_{\mathrm{DE},i}$ derived from Eq.~(\ref{eq:w}) with $H_0=67.0\pm2.0~\mathrm{km~s^{-1}Mpc^{-1}}$, $\Omega_\mathrm{m0}=0.347\pm0.03$ and the constraints on $H_i$ and $q_i$ from the P+1690 sample plus $H(z)$ data. 
		\label{fig8}
	}
\end{figure}

Figure~\ref{fig8} shows the  values of $w_{\mathrm{DE},i}$ after setting $H_0=67.0\pm2.0~\mathrm{km~s^{-1}Mpc^{-1}}$ and $\Omega_\mathrm{m0}=0.347\pm0.03$. Except for $w_{\mathrm{DE},i}$ at $z_i=0.32$, all values of $w_{\mathrm{DE},i}$ align within a 2$\sigma$ CL of $w_\mathrm{DE}=-1$.
These results are similar to those obtained using the P+1690 sample only. 

Moreover, we find that the values of $M_B$ obtained from both the Pantheon+ and P+1690 samples seem to decrease with the increase of redshift, as shown in the Table~\ref{tab4}.
To illustrate clearly  this trend, we use a simple linear function: $M_B(z)=M_0+\alpha z$  to fit the evolution of $M_B$, and obtain $M_0=-19.31\pm0.11~\mathrm{mag}$ and $\alpha=-0.64\pm 0.44~\mathrm{mag}$ for the Pantheon+ sample, and $M_0=-19.05\pm 0.15~\mathrm{mag}$ and $\alpha=-1.21\pm 0.52~\mathrm{mag}$ for the  P+1690 sample.
Both $M_0$  are consistent with that obtained from the Cepheid host~($-19.253\pm0.027~\mathrm{mag}$)~\citep{Riess2022} within 2$\sigma$ CL. 
However, the slope $\alpha$ deviates from zero by more than 2$\sigma$ CL for the P+1690 sample. Note that a decreasing $M_B$ with  increasing redshift has also been observed in the $\Lambda$CDM model when combining SN Ia data with Hubble parameter measurements and baryonic acoustic oscillation data ~\citep{Krishnan2020}. 
This trend may be attributed to various astrophysical mechanisms \citep{Hicken2009,Maoz2010,Kang2020}, or it could suggest that the expansion of the universe does not strictly follow the average evolution predicted by the FLRW model \citep{Anupama2023,Akarsu2024}. 
Furthermore, due to the degeneracy between Hubble constant $H_0$ and $M_B$, this trend could also be influenced by variations in $H_0$, as observed in several studies \citep{Wong2020,Dainotti2021,Hu2022}.

\begin{table}
	\centering
	\caption{ Constraints on $M_B$ obtained from the Pantheon+ and P+1690 samples plus the $H(z)$ data. }
	\label{tab4}
	\begin{tabular}{ccc}
	\hline
	\hline
	$z_i$ & Pantheon+ Sample & P+1690 Sample \\
	\hline
	$0.12$ & $-19.40\pm0.08$ & $-19.14\pm 0.11$ \\
	$0.22$ & $-19.40\pm0.08$ & $-19.20\pm 0.01$ \\
	$0.32$ & $-19.52\pm0.16$ & $-19.72\pm 0.18$ \\
	$0.42$ & $-19.63\pm0.13$ & $-19.55\pm 0.15$ \\
	$0.52$ & $-19.65\pm0.17$ & $-19.54\pm 0.18$ \\
	\hline
	\end{tabular}
\end{table}

\section{Conclusions}\label{sec:4}

To determine whether the cosmic evolution is consistent with the predictions of the $\Lambda$CDM model, we establish  a new and cosmological-model-independent method to explore the cosmic dynamics from observational data. 
Using the Pantheon+ sample and a model-independent P+1690 SN Ia sample, we obtain the values of $d_{L,i}$, $H_i$ and $q_i$ at five different redshift points, and calculate the EoS of dark energy $w_{\mathrm{DE},i}$ at these redshifts. 
We find that all results obtained from the Pantheon+ sample are consistent with  the predictions of the $\Lambda$CDM model within 2$\sigma$ CL.
However, the constraints on $H_i$ from the P+1690 sample deviate from the predictions of the $\Lambda$CDM model at $z_i=0.32$, $0.42$, and $0.52$ by more than 2$\sigma$ CL, with the largest deviation reaching about $3\sigma$ CL.
A similar deviation is also observed in the result for $q_i$ at $z_i=0.32$.
Moreover, the EoS of dark energy obtained from the P+1690 sample also deviates from the $-1$ line by about 3$\sigma$ CL.
After further considering the Hubble parameter measurements, we find that $q_i$ and $w_{\mathrm{DE},i}$ still deviate from the predictions of the $\Lambda$CDM model at $z_i=0.32$ by about 3$\sigma$ CL, although the constraints on $H_i$  become consistent with the model within 2$\sigma$ CL. 
We also find that a linearly decreasing absolute magnitude of SN~Ia with the increase of redshift is favored, as the slop of the linear function deviates from  zero by more than 1$\sigma$ CL.
Our results show that the $\Lambda$CDM model remains compatible with the Pantheon+ SN Ia and the Hubble parameter measurements within 2$\sigma$ CL, but only within 3$\sigma$ CL for the P+1690 sample.

\section*{ACKNOWLEDGMENTS}
This work was supported in part by the NSFC under Grant Nos. 12275080 and 12075084 and the innovative research group of Hunan Province under Grant No. 2024JJ1006.

\section*{DATA AVAILABILITY}
Data are available at the following references:
the Pantheon+ SNe Ia sample from \cite{Scolnic2022},
the P+1690 sample from \cite{Lane203},
and the latest $H(z)$ data obtained with the CC method from \cite{Cao2022}.

\appendix
\section{Checking the reliability of our method}\label{appendix}
To determine the impact of the choice of different $\Delta z$ on the constraints of the parameters $d_{L,i}$, $H_i$, and $q_i$ in Sec.~II, we plan to simulate the SN~Ia data to constrain these parameters.
We firstly employ the Kernel Density Estimate (KDE) with a band width $b=0.01$ to describe the redshift distribution of the Pantheon+ sample, and use this redshift distribution to sample  randomly 1590 points in the redshift region over $0.01<z\leq2.26$, which is the same as that  of the Pantheon+ sample.  At each redshift point, the value of theoritical apparent magnitude $\langle m_\mathrm{th}\rangle$ can be calculated from $m_\mathrm{th}=\mu_\mathrm{th}+M_B$ by assuming a fiducial model: the flat $\Lambda$CDM model with $\Omega_\mathrm{m0}=0.33$, $H_0=73.2~ \mathrm{km~s^{-1}Mpc^{-1}}$, and $M_B=-19.253~\mathrm{mag}$, which are obtained from the Pantheon+ sample.
Then, the mock $m_\mathrm{sim}$ can be sampled from the Gaussian distribution $\mathcal{N}(\langle m_\mathrm{th}\rangle,\sigma_\mathrm{SN})$. Here $\sigma_\mathrm{SN}$ is the uncertainty of  $\langle m_\mathrm{th}\rangle$, which is obtained  from the Pantheon+ sample. 
From these mock SN~Ia data, the parameters ($d_{L,i}$, $H_i$, and $q_i$) can be estimated by using the minimum $\chi^2$ method (equation~(\ref{eq:chi})),  and  $\Delta d_{L,i}\equiv d_{L,i}-d_{L,\mathrm{th}}$, $\Delta H_i\equiv H_i-H_{\mathrm{th}}$ and $\Delta q_i\equiv q_i-q_{\mathrm{th}}$ at each redshift point can also be calculated, where the subscript `th' denotes the prediction from the fiducial model.
After repeating above process 1000 times, we plot the distributions of the deviations $\Delta d_{L,i}$, $\Delta H_i$, and $\Delta q_i$.
If these 1000 deviations are concentrated around zero line, it implies that our method does not introduce any unknown errors and is reliable.

\begin{figure*}
	\includegraphics[width=0.45\textwidth]{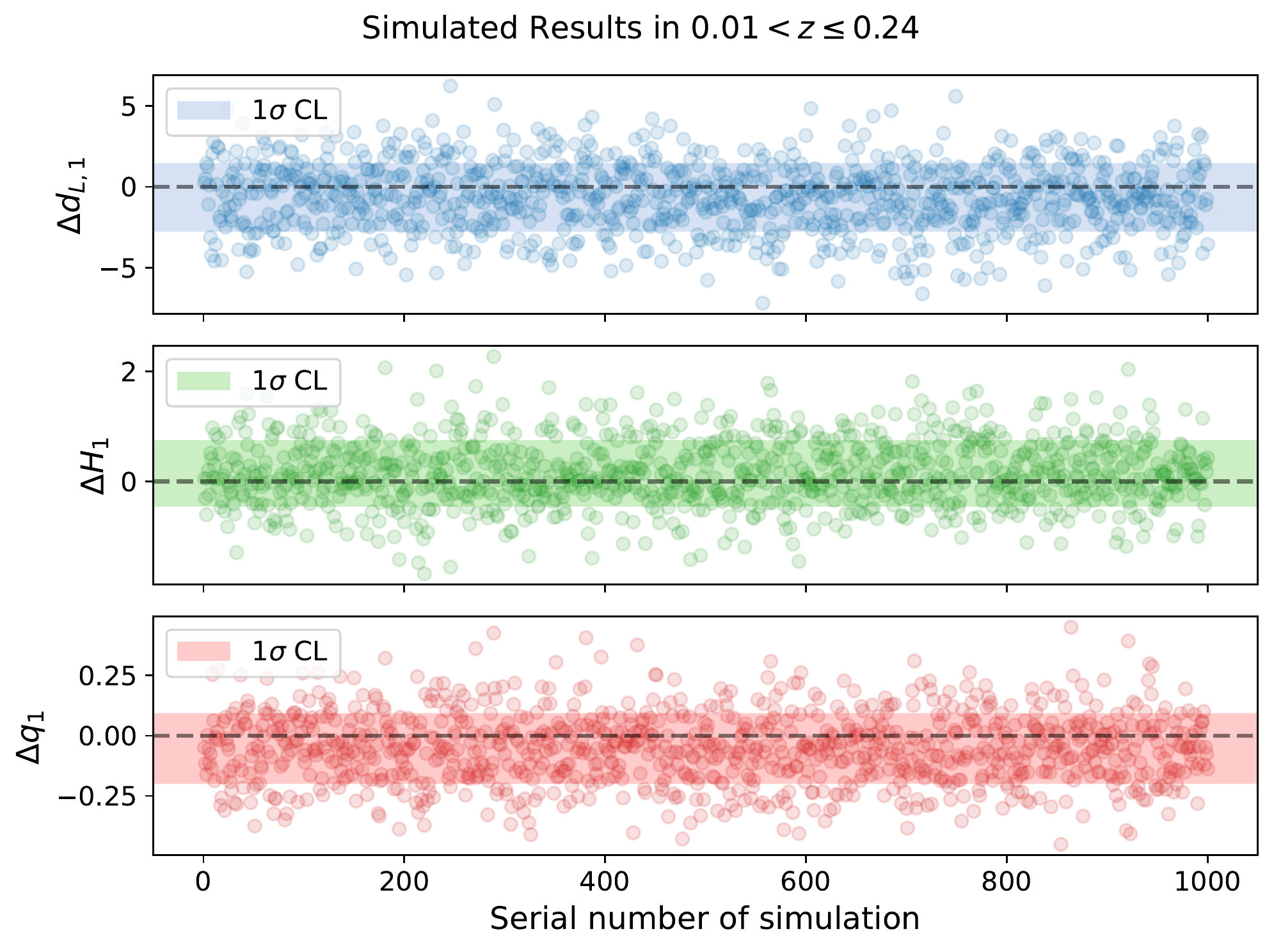}
	\includegraphics[width=0.45\textwidth]{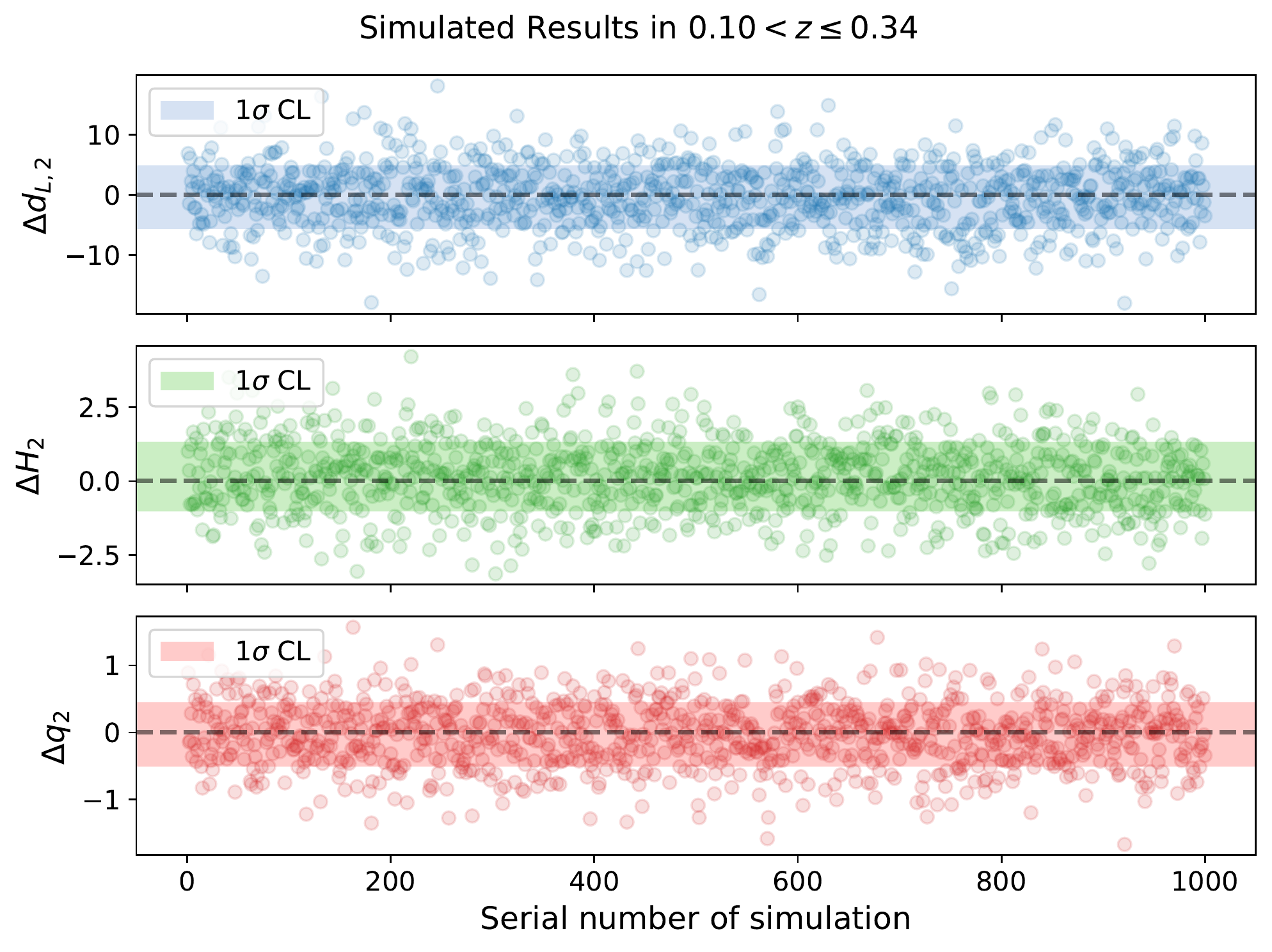}
	\includegraphics[width=0.45\textwidth]{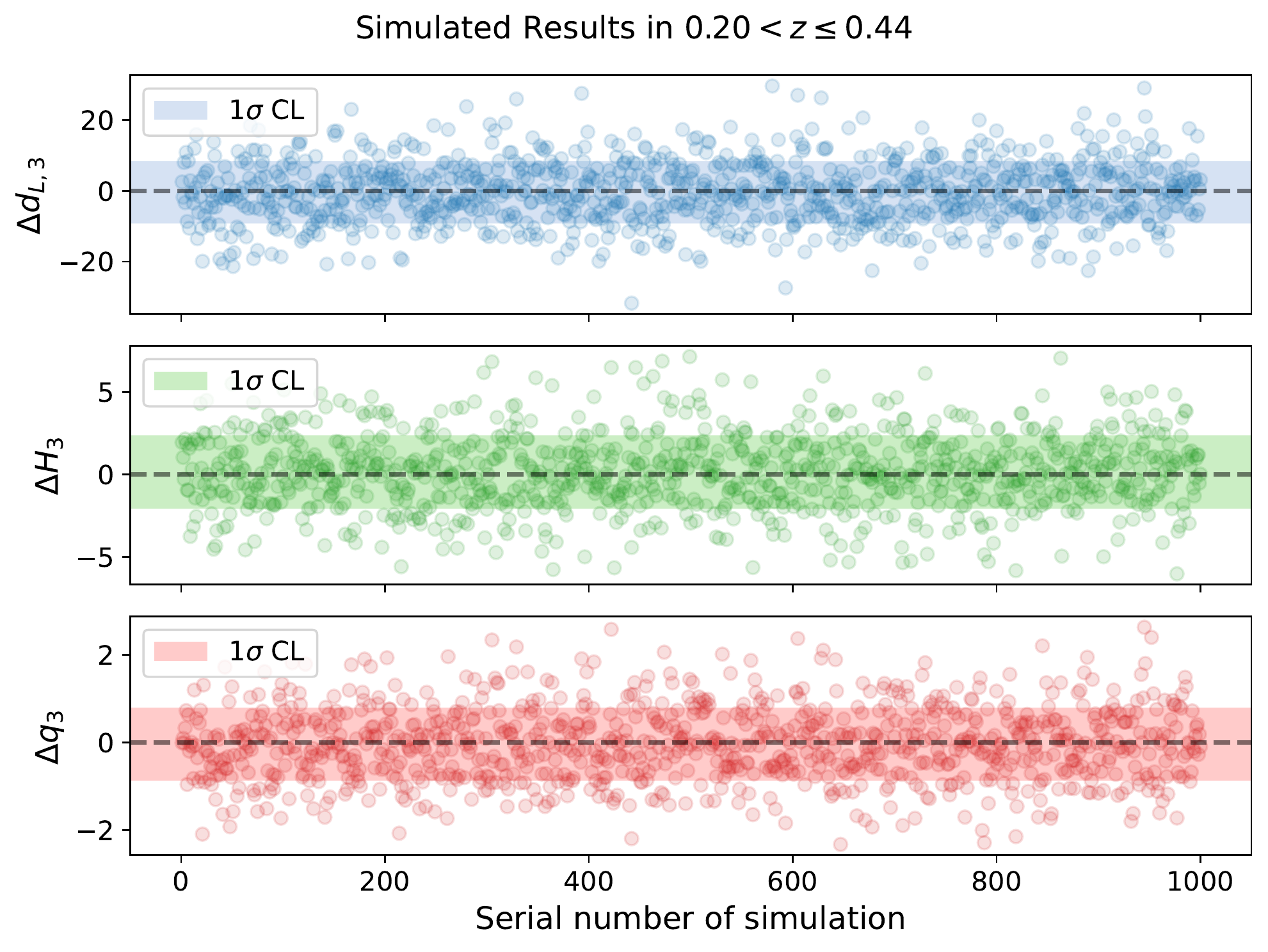}
	\includegraphics[width=0.45\textwidth]{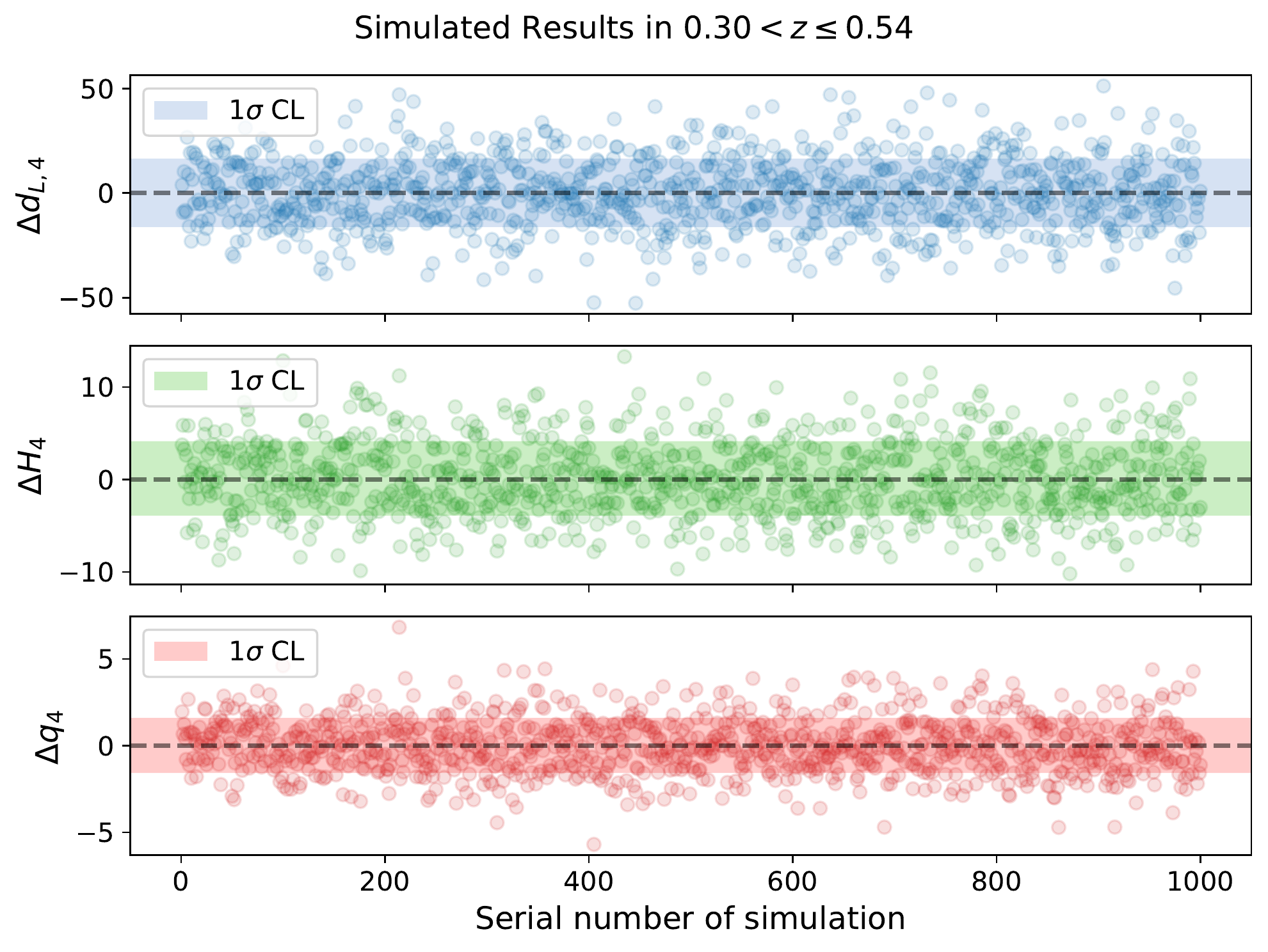}
	\includegraphics[width=0.45\textwidth]{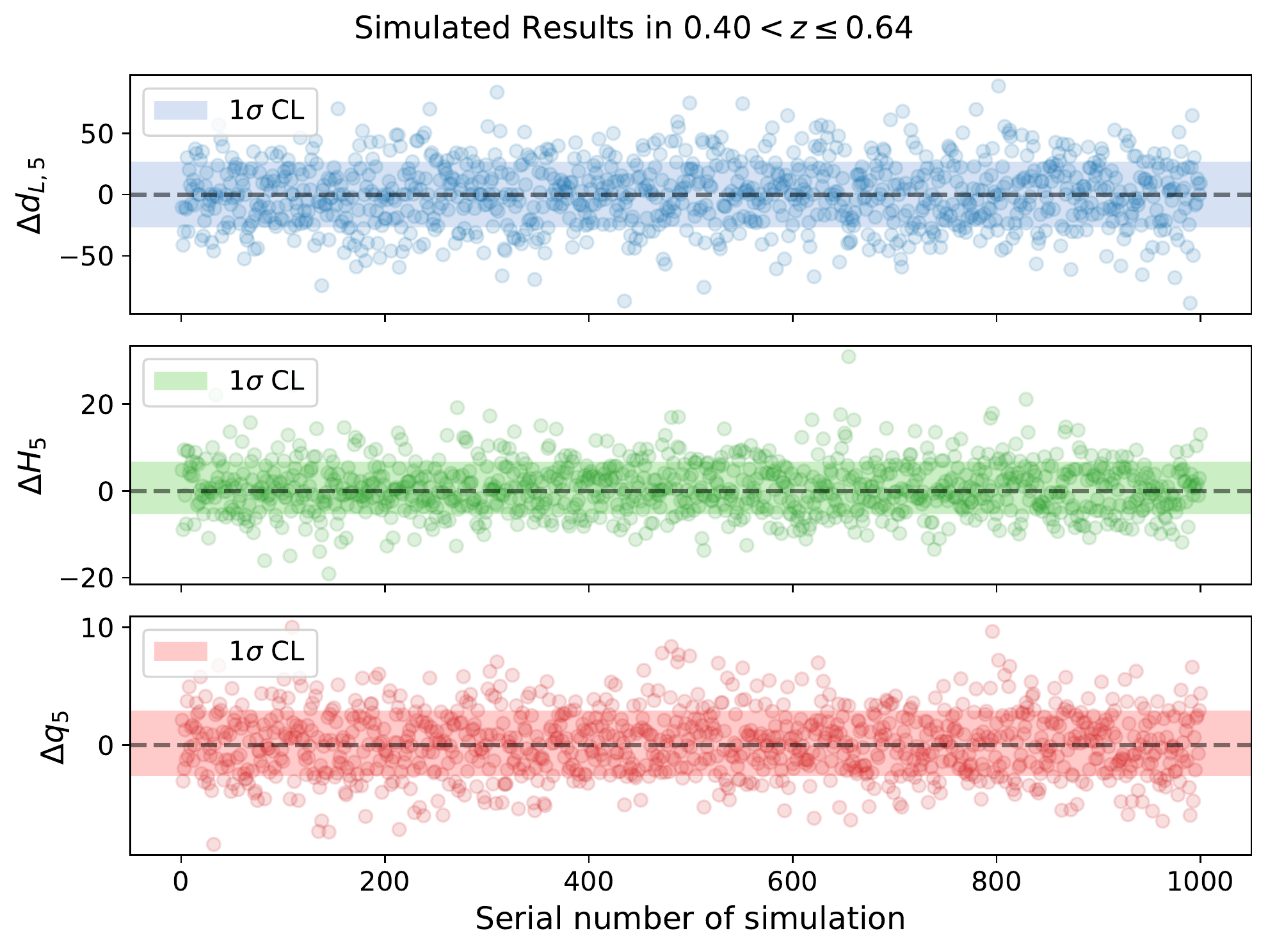}
	\caption{
		The distribution of $\Delta d_{L,i}$, $\Delta H_i$, and $\Delta q_i$ from 1000 time simulated data in the case of $\Delta z=0.12$.
		The shadow shows the $1\sigma$ uncertainty.
		\label{figA1}
	}
\end{figure*}

\begin{figure*}
	\includegraphics[width=0.45\textwidth]{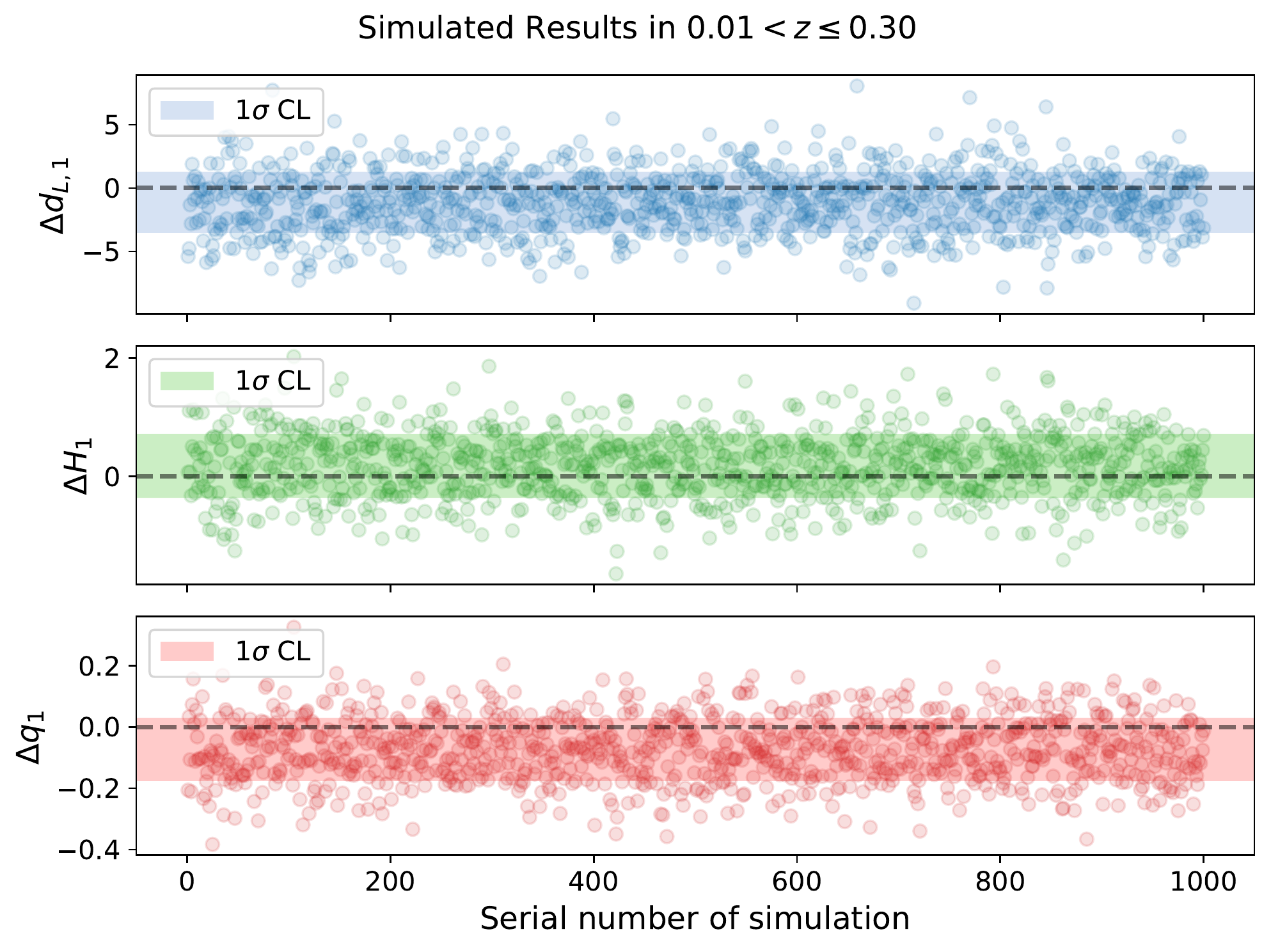}
	\includegraphics[width=0.45\textwidth]{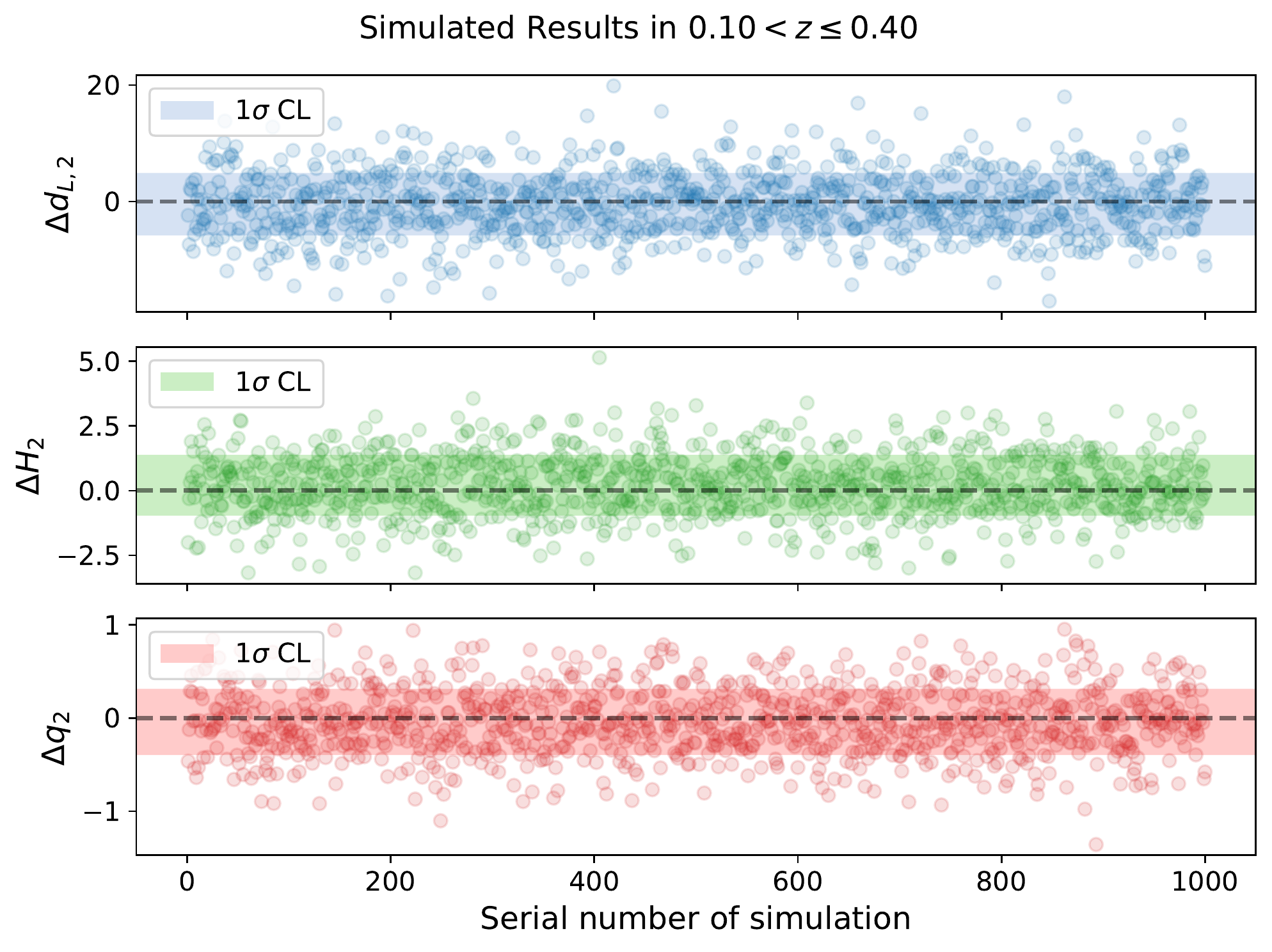}
	\includegraphics[width=0.45\textwidth]{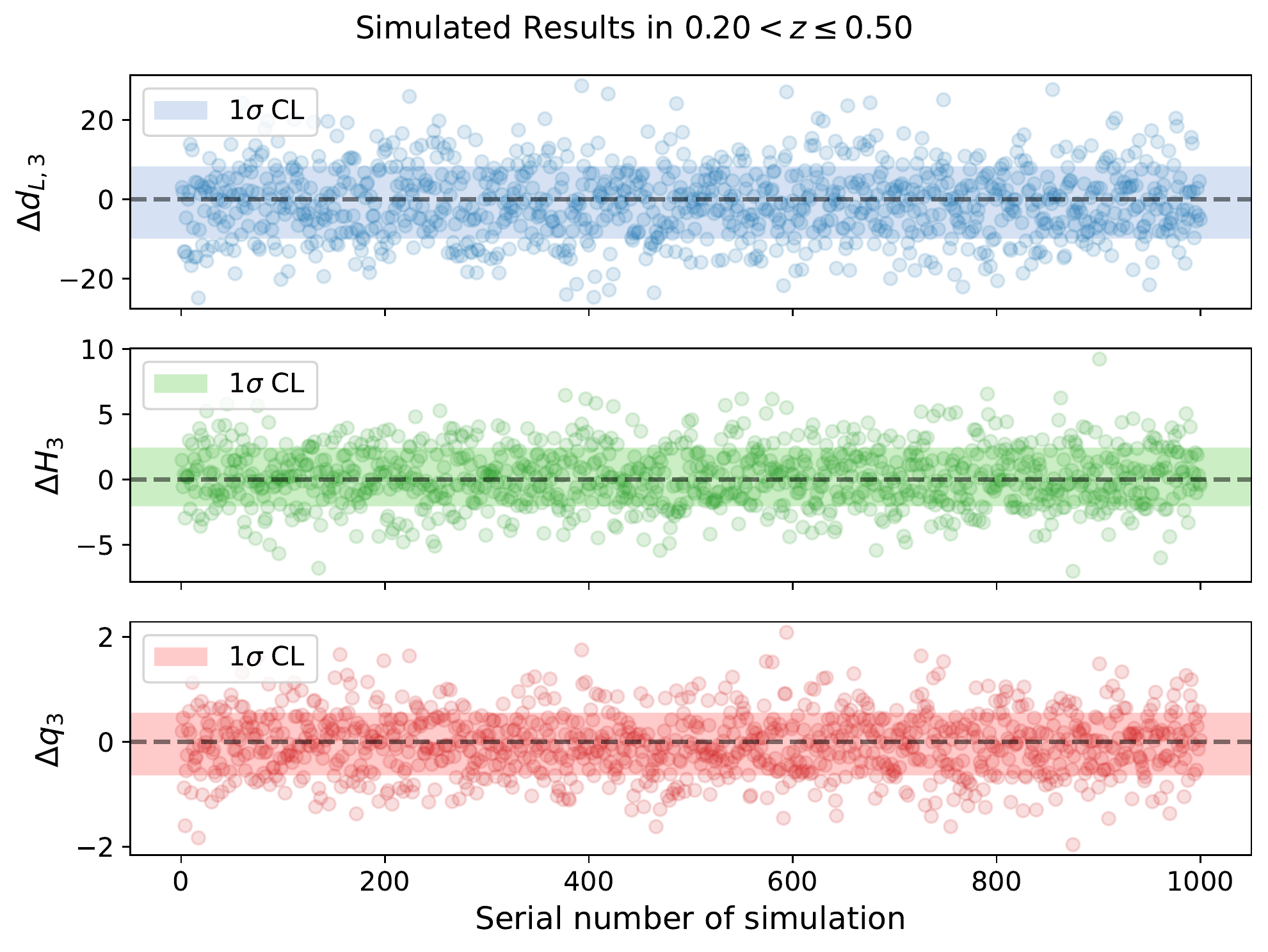}
	\includegraphics[width=0.45\textwidth]{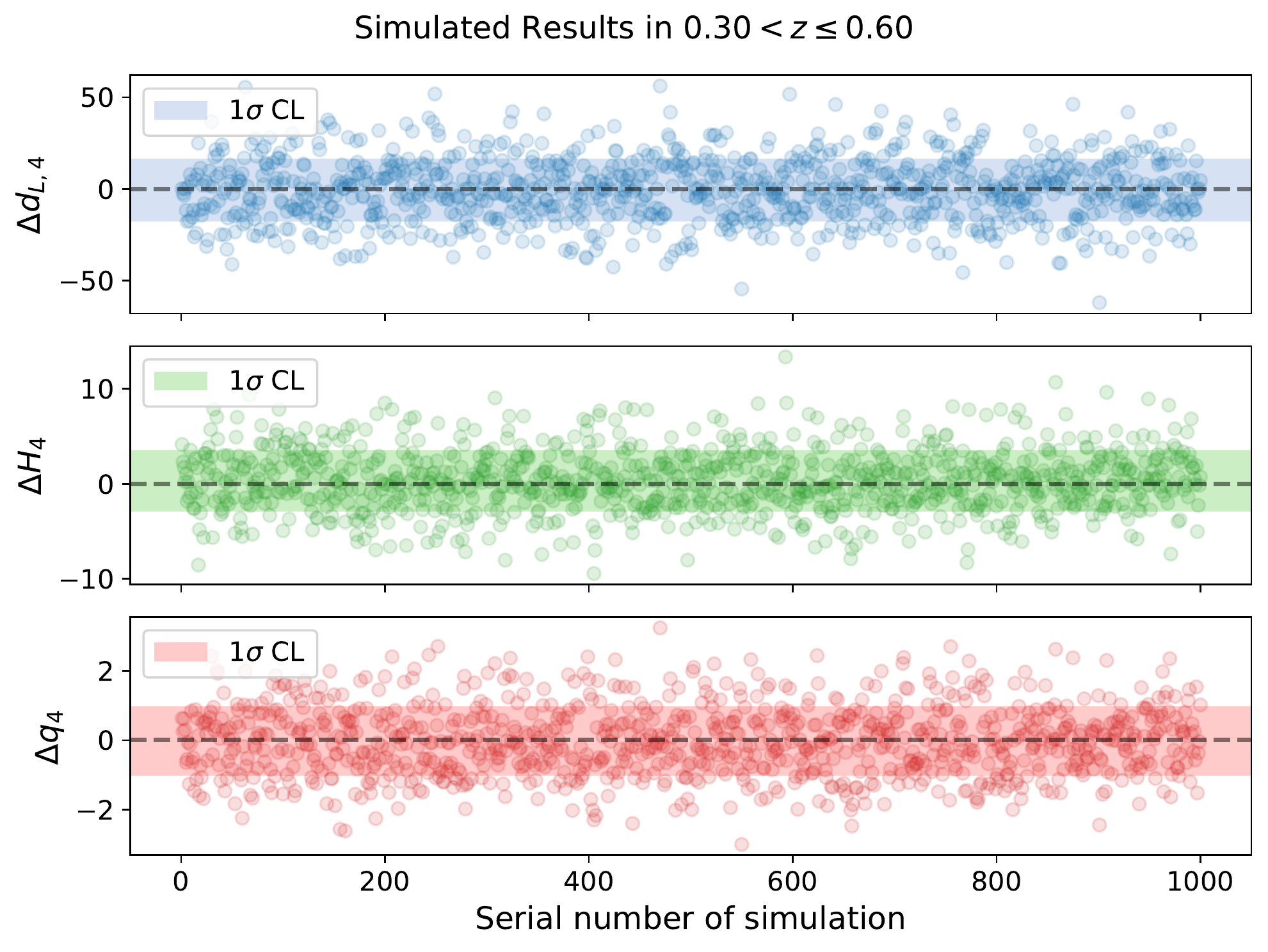}
	\includegraphics[width=0.45\textwidth]{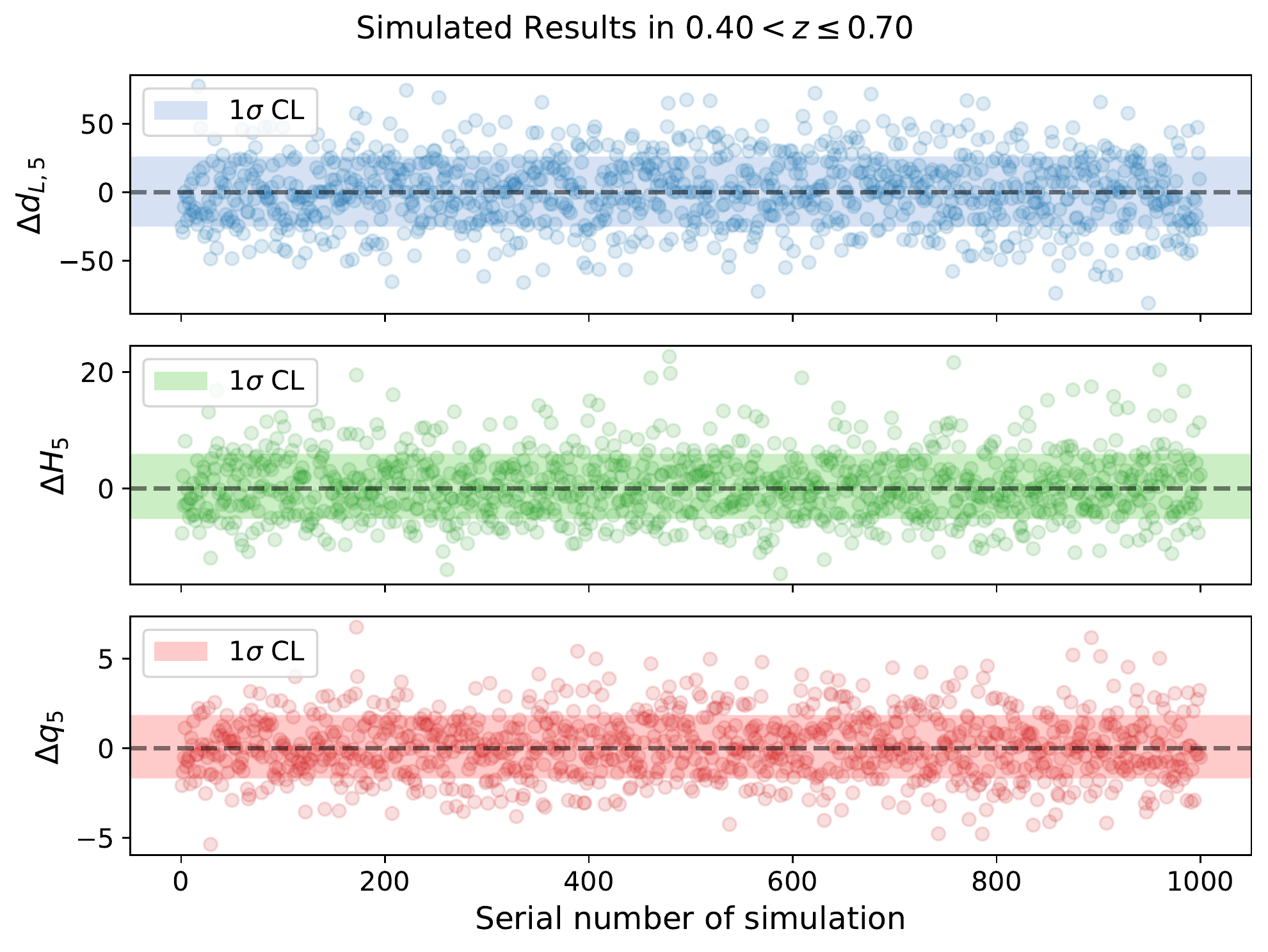}
	\caption{
		The distribution of $\Delta d_{L,i}$, $\Delta H_i$, and $\Delta q_i$ from 1000 time simulated data in the case of $\Delta z=0.15$.
		The shadow shows the $1\sigma$  uncertainty.
		\label{figA2}
	}
\end{figure*}

Here we consider two different cases: choosing  $\Delta z=0.12$ and $\Delta z=0.15$, respectively. The  redshift expansion points $z_i$ are $(0.12,~0.22,~0.32,~0.42,~0.52)$  and $(0.15,~0.25,~0.35,~0.45,~0.55)$ for  the first and second cases, respectively.
The redshift points at $z_i > 0.6$ are also calculated but the results are not shown in the paper since the constraints on the cosmological parameters are weak due to that the data points in the $\Delta z$ range are too few.
Fig.~\ref{figA1} and \ref{figA2} show the results of two cases with different redshift points, respectively.
For both cases, it is easy to see that all results are compatible with the fiducial model at the $1\sigma$ CL.
However,  in the case of $\Delta z=0.15$,  $\Delta q_1=0$ is consistent with the mean of 1000 $\Delta q_1$ only at the margin of  1$\sigma$ confidence level. 
Thus, we  choose  $\Delta z=0.12$  in our analysis.

\end{document}